\documentclass[aps,pre,amsmath,amssymb,twocolumn,showpacs,superscriptaddress]{revtex4}
\usepackage{graphicx}
\newcommand{\on}[2]{\mathop{\null#2}\limits^{#1}} 
\newcommand{\oover}[1]{\on{\circ}{#1}} 
\begin{document}
%\draft
%\twocolumn[
%\hsize\textwidth\columnwidth\hsize\csname @twocolumnfalse\endcsname
%\draft
\title{Arrested phase separation in a short-ranged attractive colloidal system: A numerical study}

\author{G.~Foffi}\email{giuseppe.foffi@epfl.ch}\affiliation{
         Institut Romand de Recherche Num\'erique en Physique des
         Mat\'eriaux IRRMA, PPH-Ecublens, CH-105 Lausanne,
         Suisse}\affiliation{ {Dipartimento di Fisica and INFM,
         Universit\`a di Roma {\em La Sapienza}, P.le A. Moro 2, 00185
         Roma, Italy} }

\author{C.~De Michele}\affiliation{ 
         {Dipartimento di Fisica and INFM,
         Universit\`a di Roma {\em La Sapienza}, P.le A. Moro 2, 00185 Roma, Italy}
        }\affiliation{
        {INFM - CRS Soft,  Universit\`a di Roma {\em La Sapienza}, P.le A. Moro 2, 00185 Roma, Italy}
        }

\author{F.~Sciortino} \affiliation{
         {Dipartimento di Fisica and INFM,
         Universit\`a di Roma {\em La Sapienza}, P.le A. Moro 2, 00185 Roma, Italy}
        }\affiliation{
        {INFM - CRS Soft,  Universit\`a di Roma {\em La Sapienza}, P.le A. Moro 2, 00185 Roma, Italy}
        }
\author{P.~Tartaglia} \affiliation{
         {Dipartimento di Fisica and INFM,
         Universit\`a di Roma {\em La Sapienza}, P.le A. Moro 2, 00185 Roma, Italy}
        }\affiliation{
         {INFM - CRS SMC,  Universit\`a di Roma {\em La Sapienza}, P.le A. Moro 2, 00185 Roma, Italy}
}

%\date{\today}
\begin{abstract}
We numerically investigate the competition between phase separation
and dynamical arrest in a colloidal system interacting via a short
ranged attractive potential. Equilibrium fluid configurations are
quenched at two different temperatures below the critical temperature
and followed during their time evolution.  At the lowest studied $T$,
the phase separation process is interrupted by the formation of an
attractive glass in the dense phase. At the higher $T$, no arrest is
observed and the phase separation process proceeds endless in the
simulated time window.  The final structure of the glass retains
memory of the interrupted phase separation process in the form of a
frozen spinodal decomposition peak, whose location and amplitude is
controlled by the average packing fraction. We also discuss the time
evolution of the non ergodicity parameter, providing evidence of a
progressively decreasing localization length on increasing the packing
fraction.  Finally, we confirm that the reported results are
independent on the microscopic dynamics.

\end{abstract}

\maketitle

%%%%%%%%%%%%%%%  TEXT  %%%%%%%%%%%%%%%%

\section{Introduction}
Advances in colloidal science make it possible to realize systems of
interacting particles, with a tunable inter particle
potential~\cite{Russel1989}.  Interactions can be controlled in the
range and in the strength, expanding considerably the possibilities
found in atomic and molecular systems.  The size of the colloidal
particles allows experimentalist to study these systems with a wide
range of optical techniques, such as Dynamic Light Scattering and
Confocal Microscopy.  Among the colloidal systems which have no atomic
counterpart are the so-called short ranged attractive colloidal (SRAC)
systems, in which the inter-particle potential has a range
significantly smaller than the colloidal size. Experimentally these
system can be realized by adding to a solution of colloidal particles
a depletant agent, generally a polymer, When the radius of gyration is
smaller than the colloidal size, the large particles experience an
effective attraction whose range is related to the size of the polymer
and the intensity to their concentration~\cite{Likos2001}.\\ SRAC
systems posses an extremely rich thermodynamic and dynamic behavior
(for recent reviews see for example Refs.~\onlinecite{Anderson2002,
Cipelletti2005, Frenkel2002, Sciortino2002}). The short range
attraction affects profoundly the structure of the phase diagram. When
the range of the attraction is comparable to the particle size, the
phase diagram presents a typical structure of a van der Waals fluid,
i.e. a liquid-liquid critical point at low density and a solid-fluid
phase transition at high density. However, when the range is much
shorter than diameter of the colloidal particles, the liquid-liquid
phase separation becomes metastable and is buried inside the
fluid-solid coexistence curve.  This phenomenon, predicted
theoretically within perturbation
theory~\cite{Gast83,Tejero1994,Dijkstra1999}, has been confirmed by
simulation~\cite{Hagen1994} and
experiments~\cite{Lekkerkerker1992,Illet1995}. It is interesting to
note that critical fluctuations related to metastable critical point
favor nucleation of the crystal phase~\cite{Auer2001}.\\ Slow dynamic
properties in SRAC systems also show features not observed in atomic
systems.  When short ranged attractions are present, the fluid-glass
line (as well as its precursors, the iso-diffusivity
lines~\cite{Foffi2002c,Zaccarelli2002b}) shows a reentrant behavior in
the interaction strength-packing fraction plane.  Two different
competing mechanism for dynamic arrest are present, generating
respectively a repulsive dominated and an attractive dominated glass.
Dynamics become so complex that the usual stretched exponential decay
characterizing the usual slowing down close to a a glass transition,
cross over toward a logarithmic
decay~\cite{Fabbian1999,Bergenholtz1999,Dawson2001,Sperl2003b,Foffi2004b,Gotze2004}.
These novel phenomena, first proposed on the basis of theoretical
calculations based \cite{Fabbian1999,Bergenholtz1999} on the Mode
Coupling Theory developed by Goetze and
collaborators~\cite{Goetze1991b}, have been subsequently confirmed by
simulations~\cite{Puertas2002,Foffi2002c,Zaccarelli2002b,Sciortino2003}
and experiments~\cite{Mallamace2000,Pham2002,Eckert2002,Chen2003}.\\
An additional phenomenon which is not usually encountered in atomic
systems takes place at low packing fractions, where SRAC systems are
known to form gel, i.e. arrested space spanning
structure~\cite{Poon1998,Verhaegh1999,Segre2001,Shah2003,Sedgqick2004}.
The nature of the gel transition in short-range attractive colloidal
systems has received significant attention in recent years (for a
recent review see for example Ref.~\cite{Trappe2004}). Several routes
to the gel state have been proposed and critically examined, with a
special emphasis on the analogies and differences between glass and
gel formation. It has been shown that, for some class of potentials,
structural arrest at low packing fractions can be interpreted as a
glass phenomenon of self-assembled
clusters~\cite{Sciortino2004,coniglio,Kroy2004}.  For the case of
SRAC, evidence is building that the gel state results from an arrested
phase separation when the phase separation dynamics generate regions
of local density sufficiently large to undergo an {\it attractive}
glass
transition~\cite{Bos1996,Soga1998,Lodge1999,Arjuzon2003,Zaccarelli2004,Foffi2004cpre,Manley2005}.\\
In this manuscript we report results of Newtonian and Brownian
molecular dynamics simulation of a binary mixture of particles
interacting via a short range attractive square well. For this system,
the location of the phase separation region and of the repulsive and
attractive lines have been previously studied.  The attractive glass
line intersect the colloidal-rich colloidal-poor coexistence curve on
the colloidal rich side at an intersection temperature $T_x$, a few
per cent smaller than the critical temperature $T_c$.  We quench
initial configurations at different densities, equilibrated at a
temperature where the system is homogeneous ($T>>T_c$) to two
temperatures lower then the critical one ($T<T_c$) and study the
coarsening dynamics.  For the case of the higher final temperature,
the coarsening dynamics never arrests and the systems slowly
approached to equilibrium phase separated state characteristic of a
conserved order parameter phase separation process. For lower
temperture quench, we find that the coarsening dynamics arrests within
the simulated time, leaving the system in an out-of-equilibrium
arrested structure.  The morphology of the resulting gel is controlled
by the phase separation process at large distances (or small
wave-vectors). At length scales comparable to the particle size, the
structure resembles the one characteristic of the attractive glass.
%%%%%%%%%%%%%%%%%%%%%%%%%%%%%%%%%%%%%%%%%%%%%%%%%%%%%%%%%%%%%%%%%%%%%%%%%%%%%%%
\section{Details of the numerical simulations}
%%%%%%%%%%%%%%%%%%%%%%%%%%%%%%%%%%%%%%%%%%%%%%%%%%%%%%%%%%%%%%%%%%%%%%%%%%%%%%%

We investigate a system that has been extensively studied earlier, a
binary square well (SW) mixture~\cite{Zaccarelli2002b,Foffi2004b}.
The binary system is a $50\%$-$50\%$ mixture of $N=2000$
particles. The two species (labeled $A$ and $B$) are characterized by
a diameter ratio $\sigma_A/\sigma_B=1.2$, which effectively suppress
crystallization. Masses are chosen to be equal and unitary,
i.e. $m_a=m_b=1$. The SW interaction is defined according to:
\begin{equation}
\label{pote}
V^{\alpha,\beta}(r) = \left\{
  \begin{array}{ll}
    \infty & \mbox{ $r<\sigma_{\alpha,\beta}$}
\\
    -u_0 & \mbox{$\sigma_{\alpha,\beta}<r<\sigma_{\alpha,\beta}+\Delta_{\alpha,\beta}$}
\\
    0 & \mbox{$r>\sigma_{\alpha,\beta}+\Delta_{\alpha,\beta}$}
  \end{array} \right.
\end{equation}
where $\sigma_{\alpha,\beta}=(\sigma_\alpha+\sigma_\beta)/2$,
$\alpha,\beta=A,B$ and $\Delta_{\alpha,\beta}$ is the range of the
attraction.  We fix $\sigma_{\alpha,\beta}$ and the well-width
$\epsilon \equiv
\frac{\Delta_{\alpha,\beta}}{\Delta_{\alpha,\beta}+\sigma_{\alpha,\beta}}=0.005$. 
The chosen $\epsilon$ value is arbitrary, but representative of all SW
potential with interaction range smaller than a few percent as far as
thermodynamic and dynamics equilibrium properties are
concerned~\cite{Noro2000,Foffi2004cpre}.  For the chosen potential $T_c \approx
0.20$.  We choose $k_B=1$
and set the depth of the potential $u_0=1$. Hence $T=1$ corresponds to
a thermal energy $k_BT$ equal to the attractive well depth. The
diameter of the small specie is chosen as unity of length,
i.e. $\sigma_B=1$. Density is expressed in term of packing fraction
$\phi=(\rho_A \sigma_A^3+\rho_B \sigma_B^3)\cdot
\pi/6$, where $\rho_\alpha=N_\alpha/L^3$, $L$ being the box size and
$N_\alpha$ the number of particles of specie $\alpha$. Time is
measured in units of $\sigma_B\cdot(m/u_0)^{1/2}$.  ND has been coded
via a standard event driven algorithm, commonly used for particles
interacting with step-wise potentials~\cite{Rapaport97}.  Between
collisions, particles move along straight lines with constant
velocities. When the distance between the particles becomes equal to
the distance where the potential has a discontinuity, the velocities
of the interacting particles instantaneously change.  The algorithm
calculates the shortest collision time in the system and propagate the
trajectory from one collision to the next one.  Calculations of the
next collision time are optimized by dividing the system in small
subsystems, so that collision times are computed only between
particles in the neighboring subsystems.  BD has been implemented via
the position Langevin equation:
\begin{equation}
\dot {\mathbf r_i} (t) = \frac{D_0}{k_B T} {\mathbf f}_i(t) + {\oover{\mathbf r}}_i(t), 
\label{Eq:langevin}
\end{equation}
\noindent
coding the algorithm developed by Strating~\cite{Strating1999}.  In
Eq.~\ref{Eq:langevin} $ {\mathbf r_i} (t) $ is the position of
particle $i$, ${\mathbf f}_i(t)$ is the total force acting on the
particle, $D_0$ is the short-time (bare) diffusion coefficient, $
{\oover{\mathbf r}}_i(t) $ a random thermal noise satisfying $<
{\oover{\mathbf r}}_i(t) {\oover{\mathbf r}}_i(0)> = k_BT \delta(t)$.
In Strating's algorithm, a random velocity (extracted from a Gaussian
distribution of variance $\sqrt{k_BT/m}$) is assigned to each particle
and the system is propagated for a finite time-step $\frac{2
mD_0}{k_BT}$, according to event-driven dynamics. We chose $D_0$ such
that short time motion is diffusive over distances smaller than the
well width. \\ Initial configurations were equilibrated at $T=1$ for
densities ranging from $\phi=0.01$ to $\phi=0.50$. For each density,
the system was quenched at two different final temperatures,
$T_f=0.05<< T_x$ and $T_f\simeq 0.15 \simeq T_x$. The constant $T$ evolution
was then followed in time.  In the case of ND, the characteristic time
of the thermostat as been chosen so that velocities reach thermal
equilibrium before the system starts to rearrange its structural
degrees of freedom.

\section{Results and Discussion}

%It is perhaps interesting to stress that, in a recent paper, we have also shown that the equivalence of the virial coefficients does not imply only thermodynamical equivalence but also, under certain condition, a trivial scaling of the dynamics~\cite{Foffi2004c}.  For the value of the well width we discuss in this work, i.e. $\epsilon=0.005$, the critical point is at $T_c=0.22$ and the critical density is $\phi_c=XXX$. 
\subsection{Phase diagram}

\begin{figure}[tbh]
\includegraphics[width=.5\textwidth]{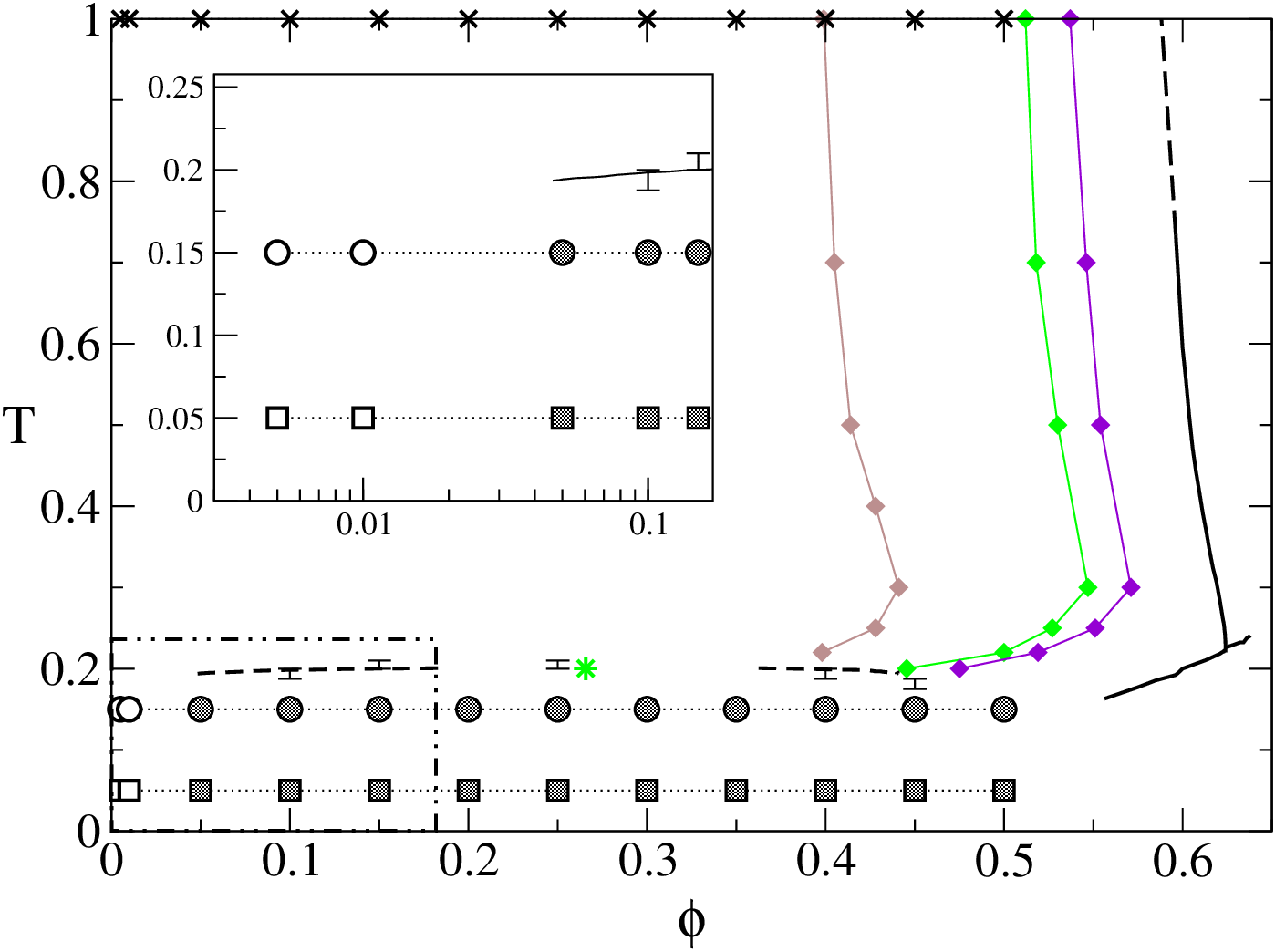}
\caption{
Graphical representation of the studied state points in the
temperature-packing fraction plane. Crosses represent the equilibrium
starting configurations at $T=1$, which, at time zero, are quenched at
$T_f=0.15$ (circles) and $T_f=0.05$ (squares) respectively.  Symbols
are shaded if, at the end of the simulation, the structure of the
system is percolating.  The error bars, calculated as explained in the
text, provide an indication of the spinodal curve for the studied
model. The present estimates are consistent with the phase coexistence
calculations of Miller and Frenkel \cite{Miller2003} for the Baxter
model (dashed line), properly transformed using the virial mapping
(Eq.\ref{bax}). The red-star indicates the location of the critical point,
from Ref. \cite{Miller2003}.  To better frame the location of the
studied state points, the Figure also shows three calculated
isodiffusivity curves, i.e. the locus of points at which the diffusion
coefficient $D/D_0=5 \times 10^{-2},1 \times 10^{-2} \mbox{ and } 5
\times 10^{-3}$ with $D_o=\sigma_B\sqrt(T/M)$.  
The bold line is the extrapolation to $D/D_0 \rightarrow0$ from
Ref.\cite{Zaccarelli2004}, rescaled to the present well width case
using Eq.\ref{bax}.  The extrapolated $D/D_o \rightarrow0$ line (see
Ref.\cite{Sciortino2003}) provides an estimate of the location of line
of dynamical arrest for this system.}
\label{fig1}
\end{figure}

The coexistence curve in the limit of vanishing well width, i.e. the
celebrated Baxter limit~\cite{Baxter1968b}, has been recently
precisely calculated by Miller and Frenkel~\cite{Miller2003}. These
data provides an accurate estimate of the coexistence curve of all
short range interacting potential, via an appropriate law of
corresponding-states~\cite{Noro2000}.  Indeed, Noro and Frenkel
noticed that, when the attraction range is much smaller than a few
percent of the particles diameter, the value of the first virial
coefficient at the critical temperature was essentially independent on
the details of the potential, down to the Baxter limit.  Following
this prescription we reported in Fig.~\ref{fig1}, the coexistence
curve calculated by Miller and Frenkel in the $\tau$-$\phi$ plane
($\tau$ is the Baxter stickiness parameter) after transforming it to
the $T$-$\phi$ plane according to
\begin{equation}
1-\frac{1}{4 \tau}=\left[1 - (e^{\beta u_0}-1)[(1-\epsilon)^{-3}-1]\right]
\label{bax}
\end{equation} 
i.e. imposing the equivalence of the virial coefficient of the Baxter
and binary SW model. To independently support the mapping, we
calculate explicitly the location of the spinodal line, by bracketing
it with the highest $T$ at which no phase separation takes place and
the lower $T$ at which we observe a growing spinodal decomposition
peak.  The calculated spinodal line, also reported in Fig.~\ref{fig1},
is in very good agreement with Miller and Frenkel estimates.\\
Fig.~\ref{fig1} also shows isodiffusivity lines for the studied SW
model, i.e. lines along which the normalized diffusion coefficient is
constant~\cite{Foffi2002c}.  These lines show the typical non monotonic
behavior characteristic of SRAC systems, in which a speed up of the
dynamics takes place at the temperature at which the competition
between the two different arrest mechanisms is balanced. The
isodiffusivity lines provide also a guide to the shape of the glass
transition line, which has been numerically estimated, by
extrapolation, as the zero isodiffusivity line~\cite{Sciortino2003}.
As previously reported, the glass line meet the coexistence and
spinodal line on the colloid-rich side, confirming that arrest in SRAC
 systems can take place at low packing fractions only as a
result of an interrupted phase separation~\cite{Zaccarelli2004, Foffi2004cpre}.\\
Fig.~\ref{fig1} also shows the location of the studied points in the phase diagram.

\subsection{Potential Energy - Number of Bonds}

\begin{figure}[tbh]
\includegraphics[width=.5\textwidth]{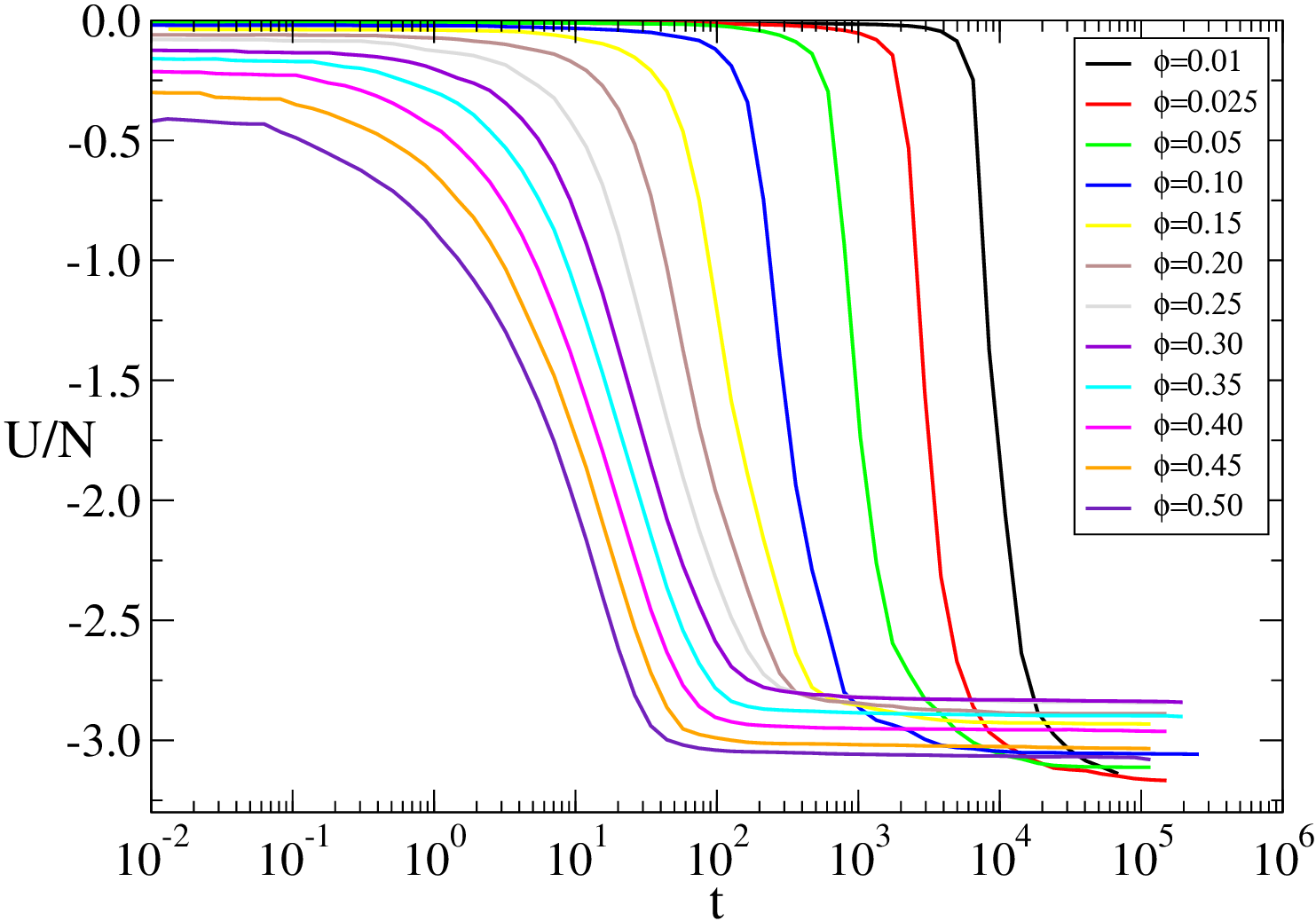}\\
\includegraphics[width=.5\textwidth]{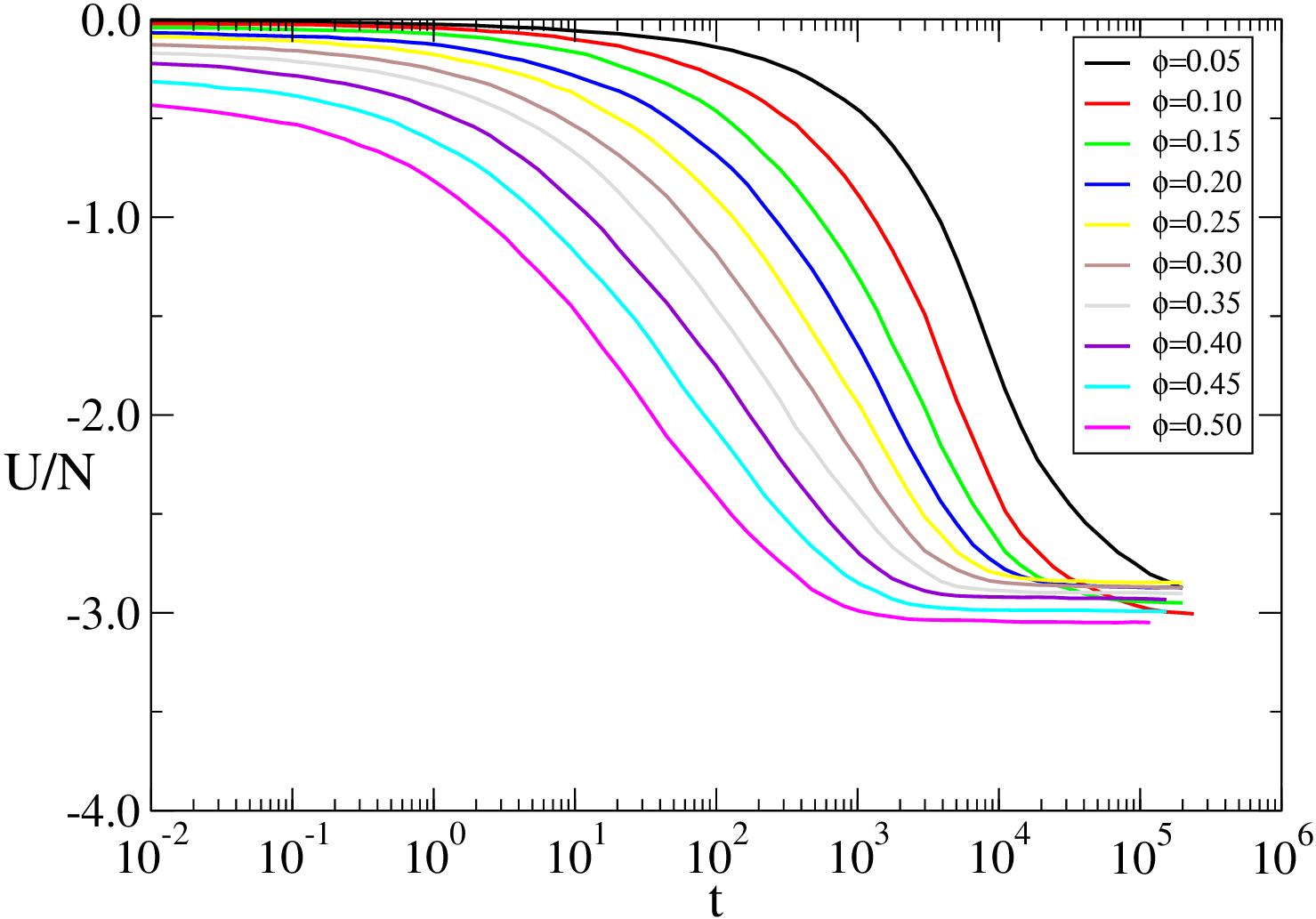}\\
\includegraphics[width=.5\textwidth]{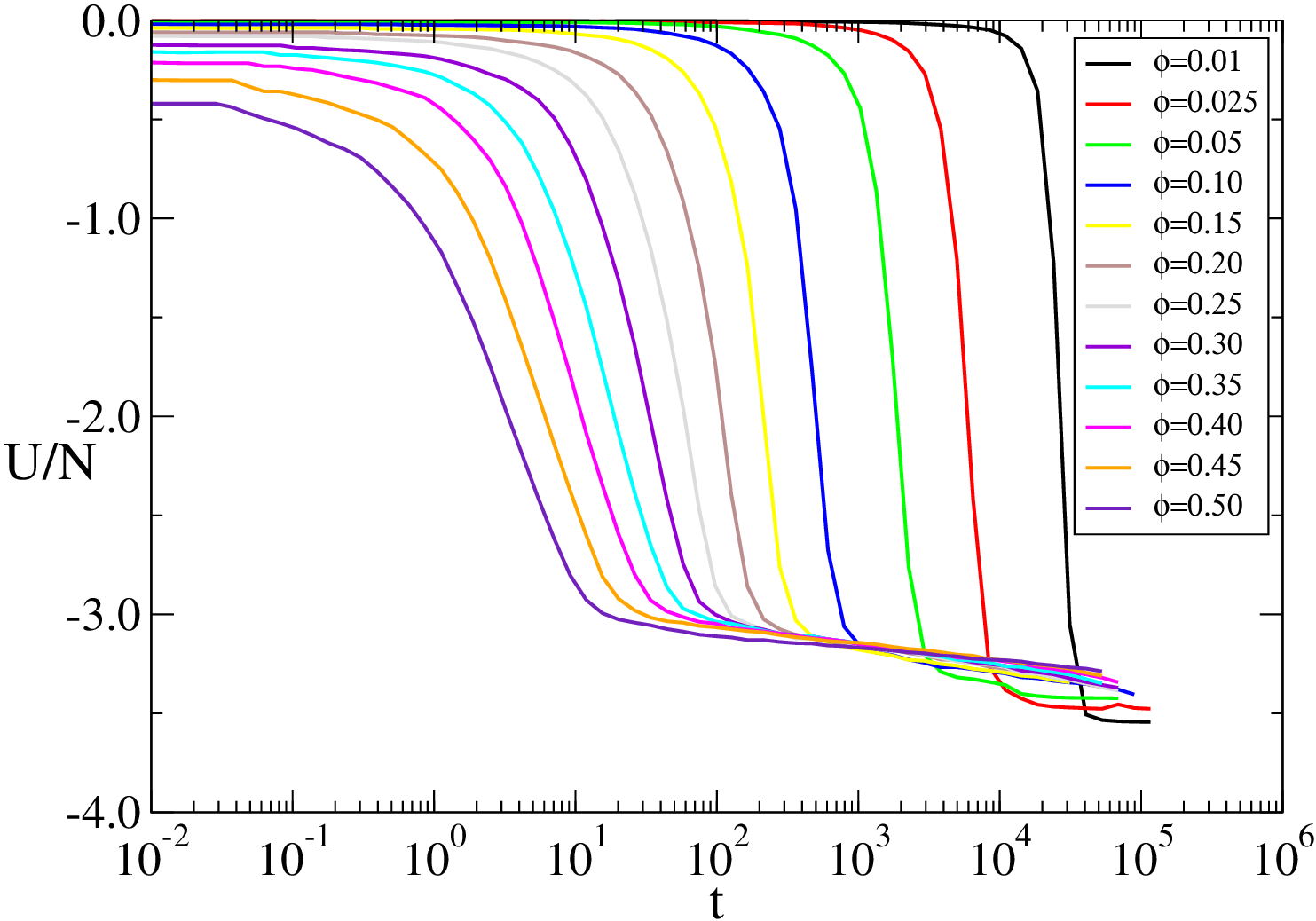}
\caption{(a) Time evolution of the potential energy per particle $U/N$  at
$T_f=0.05$, following the quench from $T=1$,  for different packing fractions.  (b) Same as (a) but for Brownian dynamics. (c) Same
as (a) but for $T_f=0.15$.}
\label{fig2}
\end{figure}

\begin{figure}[tbh]
\includegraphics[width=.5\textwidth]{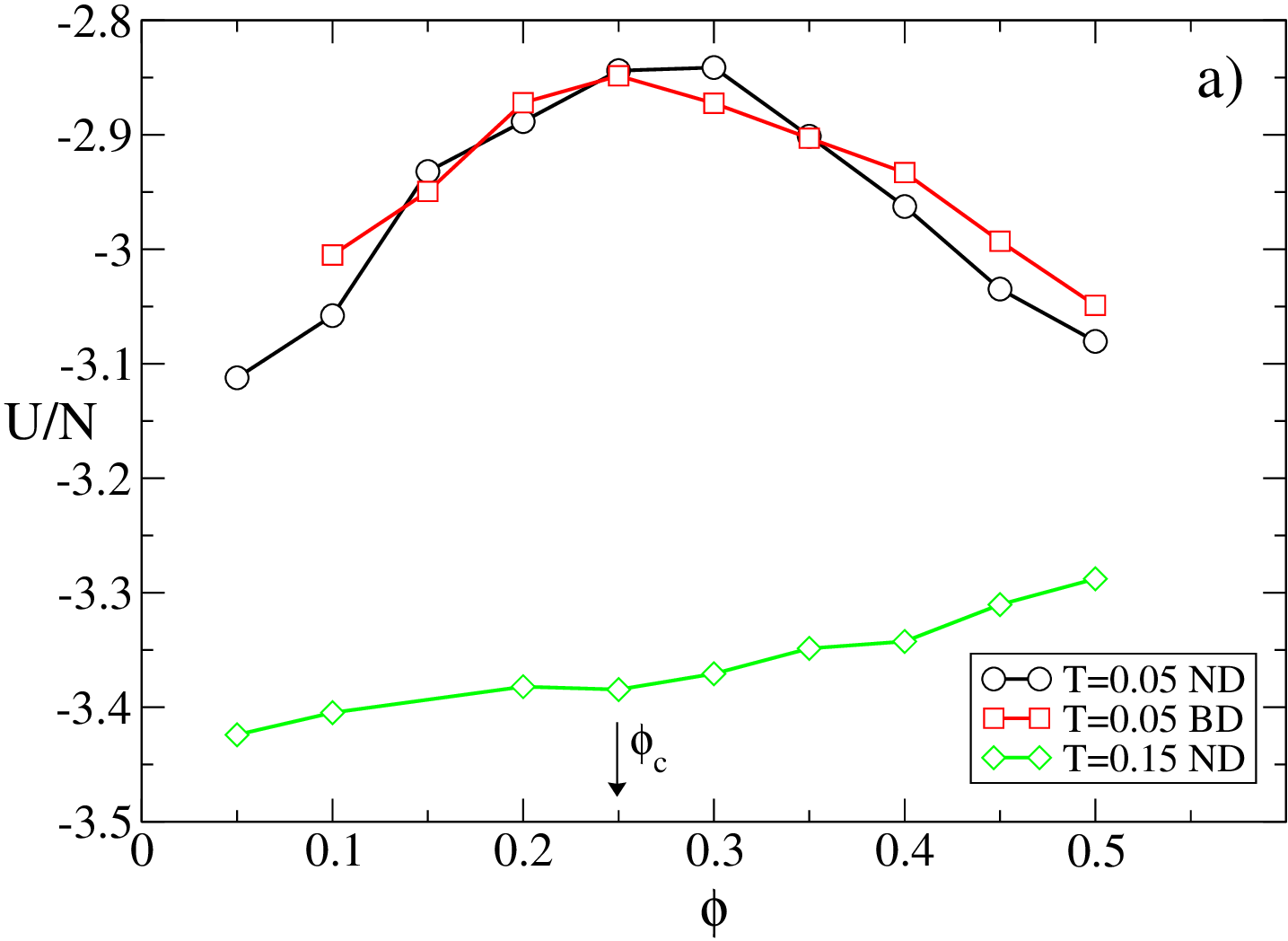}
\includegraphics[width=.5\textwidth]{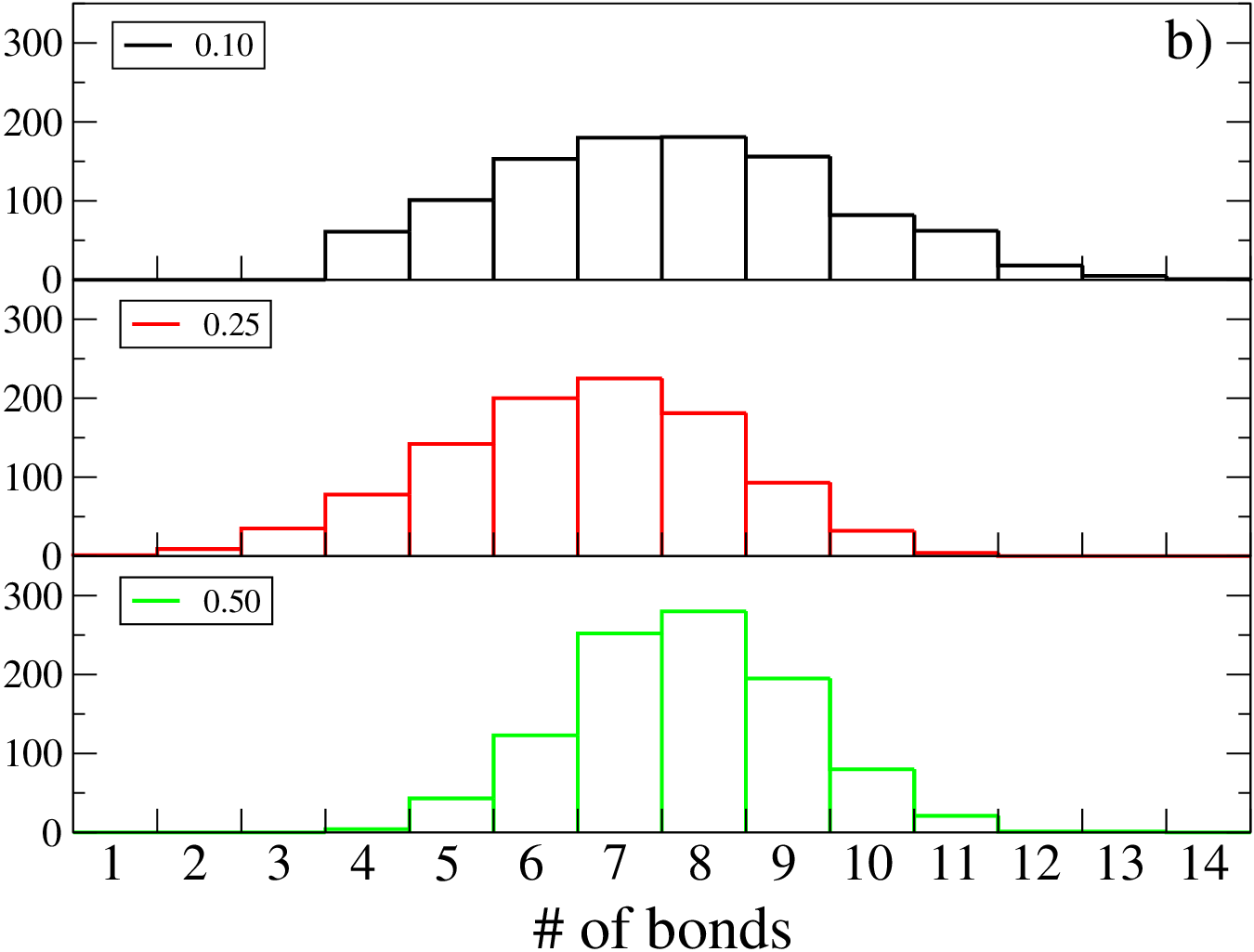}
\caption{(a) Potential energy per particle at the  end of the simulation versus $\phi$ for
$T_f=0.05$ (Newtonian and Brownian dynamics) and for $T_f=0.15$
(Newtonian). (b) Histograms of the distribution of bonds per particle
for three packing fractions: $\phi=0.10$, $\phi=0.25$, $\phi=0.50$.}
\label{fig2bis}
\end{figure}

One of the advantage of the SW models is represented by the fact that
the potential energy per particle $U/N$ is directly related to the
number of bonds $n_b$ per particle by $n_b=-2 U/(Nu_0)$. Indeed two
particle can be unambiguously considered bonded if their relative
distance is within the attractive well distance.  In the initial
configurations, i.e. $T=1.0$, the potential energy varies between
values characteristic of few particle bonded, for the lowest density,
to less then one bond per particle, for $\phi=0.50$.\\
Fig.~\ref{fig2}a reports the time evolution of $U/N$ for the ND case
and $T_f=0.05$. Following the quench at $t=0$, the velocity degrees of
freedom equilibrate and before $t\simeq1.0$ the selected $T_f$ is
reached. Structural degrees of freedom, however, relax on a much
longer time scale, dragged by the tendency of the system to phase
separate.  The evolution of the energy shows three different
processes. After a characteristic time, which is longer the smaller
the initial packing fraction, aggregation sets in and the energy
decreases significantly. This aggregation process slows down
significantly once a number of bonds of the order of 6 per particle is
reached. In this late stage of the simulations run, two different
behaviors can be distinguished. At very low density, i.e
. $\phi<0.05$, the energy continues to drop and does not seems to reach
any stationary value in the simulation time window. For $\phi >0.05$ ,
however, the time dependence of the energy abruptly stops and the
system does not show sign of further evolution. This can be seen as
first indication that an arrested structure has formed. \\
Fig.~\ref{fig2}b reports data similar to the one presented in
Fig.~\ref{fig2}a but for BD. Despite the different microscopic
dynamic, the time dependence of the energy shows similar trends and
the value of the energy at which the system stops for different
densities is indeed very similar to the ND case (see
Fig.~\ref{fig2bis}).  Comparing in more details the two dynamics, one
notice that the Brownian aggregation dynamics is smoother as compared
to the Newtonian one, probably because in ND momentum conservation
law require many body interactions for cluster aggregation.  Two body
interactions between two monomers can not produce a bounded dimer
state.  Once a small number of small clusters are present in the
system, aggregation speeds up significantly.\\ Fig.~\ref{fig2}c
reports, for ND, the time dependence of the energy for the quench at
$T_f=0.15$. Comparing Fig.~\ref{fig2}a and Fig.~\ref{fig2}c, we notice
that the major difference with the $T_f=0.05$ case is in the long time
scale. In the present case, instead of remaining frozen, the energy
continues to drift for all densities we considered (between
$\phi=0.05$ and $\phi=0.50$) indicating that the separation process
does not arrest during the simulated time.
%The $\phi$ dependence of the potential energy at the end of the
%simulation is reported in Fig.~\ref{fig2bis}. The final $U$ values are
%smaller than the ones at $T_f=0.05$, providing further evidence that
%at low $T$ a dynamic arrest transition has taken place, blocking the
%approach to an equilibrium phase separated state.  Indeed, if
%equilibrium would have been reached, $U(T=0.05)$ should have been
%lower than $U(T=0.15)$.

The $\phi$-dependence of the final values of the energy is shown in
Fig.~\ref{fig2bis}a for the three cases discussed above. For the
$T_f=0.05$ quench, on the overall range of densities this quantity
varies between $3.1$ and $2.8$ and it presents a maximum close to the
critical packing fraction. In equilibrium conditions, one would expect
the energy to grow with the packing fraction but, in this
out-of-equilibrium situation, this is note the case. To understand how
the maximum arises, we show in Fig.~\ref{fig2bis}b the distribution of
bonds for three representative packing fraction, $0.10$, $0.25$,
$0.50$. While the high and the low density cases have roughly the same
average number of bonds, at the critical packing fraction such an
average is lower, as indeed expected by the value of the energy. This
difference is originated by the presence of a large tail of particles
with low number of bonds. In other words, at the critical packing
fraction the aggregate, possess more surface particles than at lower
and higher packing. This phenomenon may results from the fact that at
the critical packing fraction critical fluctuations are stronger then
at lower and higher packing.\\ It is interesting to note that, within
numerical error, ND and BD results for $T_f=0.05$ are in
agreement. Indeed this is not a trivial issue, since one would expect
microscopic dynamics to play an important role in an out of
equilibrium situation. \\ For $T_f=0.15$, the final $U$ values are
smaller than the ones at $T_f=0.05$, providing further evidence that
at low $T$ a dynamic arrest transition has taken place, blocking the
approach to an equilibrium phase separated state.  Indeed, if
equilibrium would have been reached, $U(T=0.05)$ should have been
lower than $U(T=0.15)$.

\subsection{Structure of the aggregates in real space}
\label{RS}

An insight of the evolution of the structure during the phase
separation process can be gained by examining Fig.~\ref{fig8}, where
snapshots of the $\phi=0.10 $ system are presented for different
times, for $T_f=0.05$ (both for ND and BD) as well as for
$T_f=0.15$. Before structural rearrangements start to take place (top
row), the particles are homogeneously distributed within the
simulation box. In the second set of snapshot, the phase separation
process starts to create regions richer and lower in colloidal
particle concentration (second row), which progressively coarsen
(third row). In the final configurations (forth row), both the ND and
BD results present a very similar ramified structure for the
$T_f=0.05$ quench whilst, as expected, the structure for $T_f=0.15$ is
more compact. Comparing these intermediate states with the last
recorded one, it appears that for $T_f=0.05$ the system dynamics has
arrested. For $T_f=0.15$, however, even if the system forms a
percolating space-spanning structure, its evolution is not
arrested. Indeed the system is still able to rearrange itself,
suggesting the presence of a significant mobility.\\ It is important
to stress that the arrested structures obtained by ND and BD are
indeed very similar, which provides a support to the independence of
the non-equilibrium structure to the microscopic dynamics. Thus, if
one is interested, as in the present work, to characterize the
arrested structure, it is possible to use equivalently BD or ND since
the results is insensitive to the evolution scheme chosen. Obviously,
since BD evolves toward the final structure roughly two order of
magnitude slower then ND, the use of ND offers a net gain in terms of
computational time. If one, however, is interested in the kinetic of
the process, i.e. how the gel structure forms, issues concerning the
microscopic dynamics should be properly addressed.\\ Fig.~\ref{fig7}a
and Fig.~\ref{fig7}b show snapshots of the final configurations for
packing fractions between $0.01$ and $0.10$ for the two quenches.  At
low temperature, i.e. $T_f=0.05$, and low packing fraction,
i.e. $\phi=0.01$, the system is made up of large elongated cluster
that does not span all the simulation box, i.e. does not percolate.
Percolation is not observed at any stage during the separation
process. Some monomers are still present, but their number
progressively decreases with time. In real system the large aggregates
will be eventually subject to precipitation due to the gravitational
field. At $\phi=0.05$, the final structure percolates. All particles
belong to the spanning cluster, which is extremely open and with tiny
connections.  The situation is even more evident at $\phi=0.10$. The
network that span the box has a more defined structure and
distribution of particle appears to be more uniform. This structure
clearly resemble the idea of a gel, highly inhomogeneous percolating
arrested structure. It is perhaps interesting to stress that the
percolating structure is formed during the separation process.\\ For
$T_f=0.15$, at low packing fraction, i.e. $\phi=0.01$, the final
structure is similar to the previous case, the system has formed a
single non percolating cluster. Its structure, however, is more
compact than in the previous case resembling more the shape of
spherical droplet.  At $\phi=0.05$, the situation is different. A
single cluster is present but it does not form anymore a rigid
percolating structure but rather an elongated cluster. It is only at
$\phi\geq0.10$, that the systems percolates again. The general trend
however, is that the structures are significantly more compact than
for $T_f=0.05$, in agreement with the larger number of bonds (lower
energy) observed at $T_f=0.15$ as compared to the $T_f=0.05$ case.

\subsection{Static structure factors}
To properly quantify the structure of the system we study the static
structure factor as a function of the time elapsed after the
quench. For a binary mixture the partial static structure factors,
$S_{\alpha\beta}(q,t)$ can be defined as
\begin{equation}
S_{\alpha\beta}(q,t)=\langle\varrho^*_\alpha({\bf q},t)\varrho_\beta({\bf q},t)
%/\sqrt{n_\alpha n_\beta}
\rangle
\end{equation}
where the partial density variables are defined as
$\varrho_\alpha(\vec q)=\sum_{k=1}^{N_\alpha}\exp[i {\bf q}\cdot {\bf
r}(t)_k^{(\alpha)}]/\sqrt{N}$.  At equilibrium, time translation
invariance implies $S_{\alpha\beta}(q,t)=S_{\alpha\beta}(q,0)$,
i.e. the static structure factor is a time independent function. In
our case, however, we are in an out of equilibrium condition and the
$t$ dependence must be maintained.  In particular we shall focus on
the total static structure factor defined as:
\begin{equation}
\label{totSQ}
S(q,t)=\sum_{\alpha ,\beta}^{1,2}S_{\alpha\beta}(q,t)
\end{equation}
Numerically this quantity is calculated for a given configuration at a
time $t$ by calculating the density variable $\varrho_\alpha(\vec q)$
and averaging on different $q$-vectors with the same modulus $q$ but
different orientation. Such spherical average was performed over up to
$300$ distinct $q$-vectors. Results for the ND $T_f=0.05$ case are
shown in Fig.~\ref{fig3} for three different densities. ($\phi=0.10$
in Fig.~\ref{fig3}a), critical ($\phi=0.25$ in Fig.~\ref{fig3}b) and
high ($\phi=0.50$ in Fig.~\ref{fig3}c).\\ At $\phi=0.10$, the
structure factor starts from the typical shape of a low density hard
sphere fluid, i.e. no structural peaks either at contact, i.e. $q\sim
6.2$ or at higher values. As the energy starts to drop, roughly at
$t\sim10^2$, the system becomes more structured. A first effect is
represented by an increase at the nearest neighbor peak of the
structure factor, a structural confirmation of bond
formation. Simultaneously, a spinodal decomposition peak (at low $q$)
starts to emerge. This indicates the  development of a characteristic
length roughly proportional to $2\pi/q_p$, where $q_p$ is the location
of the peak. This peak grows, coarsen and eventually stops.\\ To
characterize the evolution of the peak we studied the evolution of the
maximum and of the first moment $q_1$ of $S(q)$. This moment is
defined as:
\begin{equation}
q_1 \equiv \frac{\int_q^{q_c} q S(q,t) dq}{\int_0^{q_c} S(q,t) dq}
\end{equation}
where the integral is performed with a proper cutoff $q_c$. This
quantity scales as the peak location but can be calculated with better
accuracy~\cite{Glotzer1994,Zaccarelli2004}.  We have chosen $q_c=4.0$,
to weight the long wavelength fluctuations responsible for the phase
separation. $S(q_p,t)$ and $q_1(t)$ are shown in the inset of
Fig.~\ref{fig3}a. The maximum of the peak start to grow at $t\sim10^2$
and then at $t\sim 4 \times 10^3$ reaches a value around $20$ and it
remains frozen at this value.  This is also confirmed by the behaviour
of $q_1$ that after a transient time reaches a plateau. As discuss
before ~\cite{Soga1998,Lodge1999,Zaccarelli2004}, these are phenomena
that are typical of the formation of a dynamic arrested
structure~\cite{Segre2001,Shah2003}.\\ For density close to the
critical one, i.e. $\phi=0.25$, the behaviour is similar to the
previous case, but less pronounced, as shown in
Fig.~\ref{fig3}b. $S(q,0)$ presents a more pronounced contact peak, a
consequence of the higher density. As the system evolves in time the
contact peak grows as in the $\phi=0.10$ case. The dynamical evolution
of the maximum and of the first moment $q_1$, shown in the inset of
Fig.~\ref{fig3}b, is similar to the previous case but the final
plateau values are much lower. Hence in this situation the arrested
phase presents smaller aggregates and is more homogeneous then the
case at lower density, quantifying the previous observations based on
the real space pictures (see Sec.~\ref{RS}).\\ A different situation
emerges when the system is at high density. The evolution of the
structure factor for $\phi=0.50$ is presented in
Fig.~\ref{fig3}c. $S(q,0)$ presents a very pronounced contact peak,
higher then $2.5$ due to the high particles packing. After the quench,
when the system evolves, this first peak decreases and moves to
slightly higher $q$, i.e. the opposite trend of the previous
case. This is not surprising, it is well known that when the system at
high density an temperature is cooled the effect for SRAC systems is
the lowering of the first peak and the appearance of long tail
oscillations. Indeed in MCT this trend is responsible for the
reentrant glass line and for the formation of the attractive
glass~\cite{Dawson2001} and it has been observed in simulation on SW
systems both at equilibrium~\cite{Zaccarelli2003} and in the aging
regime~\cite{Foffi2004b}. It is interesting to note that, while the
contact peak decreases with time, the second peak ($q\sim12.5$)
increases, a clear indication of the emergence of these oscillations.
Also for lower $q$, the situation is different from the two densities
discussed above.  A slightly increase of in the structure factor is
present with time but now we can not talk of a real peak
formation. The system freezes but its structure remains basically
homogeneous and no large aggregates are formed, as confirmed by the
evolution of the maximum and of $q_1$ reported in the inset of
Fig.~\ref{fig3}c.  These results suggest that at this packing fraction
the system is sufficiently homogeneous that the MCT scenario is
recovered. The frozen structure is similar to the one of an
homogeneous liquid and the contact interactions are responsible for
the structural arrest, enforcing the idea that SRAC gels are spatially
inhomogeneous attractive glasses (in the MCT fashion), where the
inhomogeneity is build by the phase separation process.\\ Analysis of
the BD configurations confirms the above scenario, in agreement with
the previous discussion based on the potential
energy. Fig.~\ref{fig4}a, Fig.~\ref{fig4}b and Fig.~\ref{fig4}c report
BD results for $\phi=0.10$, $\phi=0.25$ and $\phi=0.50$
respectively. Results are very similar to that we previously discussed
for ND. For $\phi=0.10$ and $\phi=0.25$ a low $q$ peak emerges and get
arrested at a plateau. The difference with ND results is represented
by the time it takes to get to the maximum value of the low $q$
peak. In the BD, the transient is slower as was found in the energy
evolution. Indeed this phenomenon is more evident for $\phi=0.10$,
where in the explored simulation window, only the approach to the
plateau can be studied.  At $\phi=0.50$, no low-$q$ peak emerges in
agreement with the ND results.\\ The situation is different for the
higher temperature quench case, i.e. $T_f=0.15$. Fig.~\ref{fig5}a,
Fig.~\ref{fig5}b and Fig.~\ref{fig5}c report $S(q,t)$ for $\phi=0.10,
0.25, 0.50$ respectively. As for the $T_f=0.05$ case, following the
quench the system starts to develop inhomogeneities, expressed by the
formation of a growing low $q$ peak in $S(q,t)$. In this case,
however, no signs of structural arrest are observed. For example for
$\phi=0.10$, Fig.~\ref{fig5}a, $S(q,t)$ develops a maximum at
$t\sim10^3$ that shows a slow but continuous increase. Similarly the
position of the peak, represented by $q_1$, moves to lower $q$ and
continues to drift for all the time of the simulation, indicating that
the aggregates are growing in size. At later stage, $t\sim 4\times
10^4$, the slow growth of the maximum presents a sharp increase giving
evidence that a different regime has been reached. Similar behaviour
is observed for $\phi=0.20$, Fig.~\ref{fig5}b, but in this case the
increase in the maximum and position of the first peak is faster as
expected since the quench is deeper in the phase separating region.
Finally for $\phi=0.50$, Fig.~\ref{fig5}c, there is an evident
formation of a low $q$ peak that, differently from the $T_f=0.05$,
slowly grows.\\ To conclude our analysis, we present $S(q)$ of final
configurations at all the densities that have been investigated with
ND. Fig.~\ref{fig6}a shows the $T_f=0.05$ case. The first thing to
notice is that the shape of the structure factors does not
significantly depend on $\phi$ when $q \geq 5.0$, suggesting a similar
local structure. At lower $q$, however, differences are evident. For
the lowest density, $\phi=0.01$ and $\phi=0.05$, the system proceeds
with the phase separation but for higher density this phenomenon is
arrested. Moreover, increasing $\phi$ the degree of dis-homogeneity in
the system decreases continuously up to $\phi=0.50$ where a low $q$
peak is not detected and the structure is very similar to that of
homogeneous liquid. From this analysis it is evident that we can
distinguish between two different $\phi$ regions. (i) At low packing
fraction, i.e. $\phi<0.10$ the system does not present any arrest in
the phase separating process. (ii) In the intermediate regime, i.e
$0.10<\phi$, the system gets arrested in a structure that presents a
lower degree of dis-homogeneity as the density is increased.  The
cross-over between these two regimes is characterized by the
connectivity properties of the resulting structure. The onset of a
percolating attractive glass structure appears to be the condition for
global arrest. At small packing fractions, glass clusters can still
freely diffuse and aggregate progressively. Formation of a gel built
with a diffusion limited cluster mechanism in which the glassy
clusters created in the phase separation process are the renormalized
monomers is a possibility which can not be excluded, but it would
requires the study of a much bigger system, which is out of our
present capabilities. This issue remains open for further
investigations.\\ In the $T_f=0.15$ case, the low $q$ regime grows
indefinitely since at this value of the temperature the phase
separation is taking over and the effect of the low temperature
affects only the kinetics of this process.

\subsection{Structure of the aggregates in two dimensional slabs}

In this section we focus our attention on the properties of two
dimensional {\it slabs} of the simulation box along one of the
axis. This is something that, in general, is achieved in confocal
microscopy experiments. Indeed, the typical size of a colloidal
particle allows for this very powerful investigation in real
space. Among the important phenomena that has been studied by this
technique we can mention structural relaxation in hard sphere
mixtures~\cite{Weeks2000}, nucleation phenomena~\cite{Gasser2001},
fluid-fluid interfacial properties~\cite{Aarts2004} and indeed
aggregation and gelation
phenomenon~\cite{Kilfoil2003,Poon2004,Stradner2004}.\\ We divide our
simulation box in slabs along the $z$ axis.  %It is perhaps worth
mentioning that in our simulation there is no external field acting on
the system (such for example gravity) and consequently the choice of
the height of the slabs is arbitrary.  We take slabs of thickness
$\Delta=4$, i.e. four times the diameter of the $B$ particles. We
focus on the final configuration at $\phi=0.10$ and $T_f=0.05$, shown
in Fig.~\ref{fig9}. Fig.~\ref{fig10}a shows snapshots from the top of
three arbitrary slabs. The structure is extremely ramified and it is
possible to notice long elongated clusters that form the basic unit of
the arrested structure. As expected, the distributions of particles in
the three slabs are very similar since the phase separating process
was arrested before a big compact cluster was formed. The situation is
different for the $T_f=0.15$ case. The three slabs present very
different distributions, showing the presence of a big aggregate that
results from the tendency of the system to form a single spherical
droplet, since no arrest is taking place. On the contrary in a
different slab, only few particles are present.\\ To better
characterized this nearly two dimensional configuration, we calculate
the two dimensional structure factor $S_{2d}(q)$ defined as:
\begin{equation}
S^{2D}(q,z)=\sum_{\alpha ,\beta}^{1,2}\langle\hat\varrho^*_\alpha(\hat{\bf q}),\hat\varrho_\beta(\hat{\bf q},z)
\rangle
\end{equation}
with,
\begin{equation}
\hat\varrho_\alpha(\hat{\bf q})=\frac{1}{\sqrt{N(z)}}\sum_{k=1}^{N(z)}\exp\left[i 
\left(q_x x_k^{(\alpha)}+q_y y_k^{(\alpha)}\right)\right]
\end{equation}
where the sum is restricted to coordinates belonging to the particle
that lies within $z+\Delta/2$ and $z-\Delta/2$, $N(z)$ is the number
of such particles, $n_\alpha(z)$ the relative concentration of the two
species in the slabs. The vector $\hat{\bf q}$, lies in the plane of
the slabs and has the chosen modulus whereas the orientation is chosen
randomly.  For the total two dimensional structure factor a definition
analogous to the one given by Eq.~\ref{totSQ} holds.  Fig.\ref{fig11}a
reports the results for $T_f=0.05$ for the three $z$-values considered
together with the spherical averaged $S(q)$ considered before.  In all
cases in which the slab dimensions are representative of the sample,
the three dimensional $S(q)$ coincides with the two-dimension one.

\subsection{Non ergodicity parameter}

When a system forms an arrested state only few of the possible
configuration in phase space are actually explored, i.e. the system is
non-ergodic. This is normally detected by the density correlation
functions, $\phi^{\alpha\beta}_q(t,t_w)$ defined by:
\begin{equation}
\phi^{\alpha\beta}_q(t,t_w)=\langle\varrho^*_\alpha({\bf q},t)\varrho_\beta({\bf q},t+t_w)\rangle
\end{equation}
where configuration at a time $t_w$ after the quench is correlated
with a configuration at $tw+t$.  In what follows we focus our
attention on the total correlator defined, in a similar fashion to
Eq.~\ref{totSQ} by
\begin{equation}
\label{totPHI}
\phi_q(t, t_w)=\sum_{\alpha ,\beta}^{1,2}\phi^{\alpha\beta}_q(t,t_w)
\end{equation}
When the system is at equilibrium, time translational invariance holds
and the correlators are independent from the waiting time,
i.e. $\phi^{\alpha\beta}_q(t,t_w)=\phi^{\alpha\beta}_q(t)$. When the
system starts to lose ergodicity, correlators do not relax anymore to
zero. This effect is usually measured in terms of the non ergodicity
parameter $f_q$, defined as the long time limit of the correlator,
i.e. $f_q=\phi_q(t\rightarrow\infty)$. This quantity represents the
order parameter for the glass transition, since when it is zero the
system is in an ergodic state, when it is finite the system is in a
non-ergodic state. In equilibrium, within MCT formalism it is possible
to directly calculate $f_q$ from the static structure factor and
eventually test the theory with experimental or numerical results. We
studied both the density-density correlation function and its long
time value. We focus on the case $T_f=0.05$ and we chose to analyze
the system closer to the critical packing fraction, i.e. $\phi=0.25$\\
In Fig.~\ref{fig12}a, the evolution of the correlation function,
expressed as $\phi_q(t-t_w)$, is presented for waiting times longer
then the time needed to thermalized the velocity degree of freedom and
for a representative $q$-vector, $q\sigma_B=20$.  For short waiting
time, the correlation functions relax to zero. For larger waiting
time, a non ergodic contribution arises. Eventually, for large enough
$t_w$, the correlators remains on a plateau that is very closed to
unity. This behavior is different from the evolution of the aging
dynamics for SRAC systems at higher density~\cite{Foffi2004b}. In that
case, after a low temperature quench, there was no evident sign of a
plateau even at the largest waiting time considered. In other words,
the aging dynamics of the attractive glass is different from the
present one, a fact that enforce the idea that the two phenomena have
different origins.\\ The evolution of $f_q(t_w)$ with waiting time as
obtained from the long time limit of the correlators is shown in
Fig.~\ref{fig12}b for four representative waiting times. The
wave-vector dependence of $f_q(t_w)$ develops progressively, starting
from small $q$.  The width of the non-ergodicity parameter is related
to localization length. As time goes on, more and more particles
aggregate, resulting in a decrease of the average localization
length. The fact that already at short time the system posses a finite
low $q$ non-ergodicity parameter is a clear indication of the presence
of a spanning structure made of few particles. As more particles join
the percolating cluster the system becomes more non ergodic on shorter
and shorter length scale.  This suggests that the localization length
progressively decreases, following the same pattern recently observed
in both chemical~\cite{Saika2004} and
thermoreversible~\cite{Zaccarelli2004pre} gels. \\ We have attempted
to compare the non ergodic behavior observed in our simulations with
MCT predictions, using as input the $S(q)$ calculated from the
simulations.  Unfortunately, as well known, MCT overestimates
dynamical arrest.  In the case of HS, arrest is predicted to take
place for $\phi> 0.516$ if the Percus-Yevich $S(q)$ is used and for
$\phi> 0.525$ with Verlet-Weiss correction or, more precisely, for
$\phi > 0.546$ is the ``exact" $S(q)$ calculated from simulations is
chosen~\cite{Foffi2004}.  In both cases, these critical $\phi$ values
are smaller than the experimentally and numerically detected value of
$\phi=0.58$.  Similarly, the MCT predictions for the attractive glass
overestimate the ideal glass transition temperature by more than a
factor of two.  Comparing experimental or simulation data with MCT
predictions for the ideal glass transition locus requires an
appropriate mapping in the $\phi$-$T$
plane~\cite{Sperl2003,Sciortino2003}.  MCT predictions which do not
account for the mapping suggest that the ideal glass line preempts the
spinodal line (i.e. it is located above the phase separation
curve)~\cite{Bergenholtz2003}.  Only when the appropriate mapping is
accounted for the attractive glass line correctly ends in the high
colloid concentration side of the spinodal
curve~\cite{Zaccarelli2004}.  It is very unfortunate that the
overestimate of the glass critical line prevents the possibility of
meaningful solving the MCT equations for the case of the phase
separating system by using as input the ``exact" numerical $S(q,t_w)$,
since the role of the large $q$ is dominating already at short
time. Our attempts to solve the MCT equations failed in reproducing
the small width observed in $f_q$ at short times ($t=57$ in
Fig.~\ref{fig12}b). The first non vanishing $f_q$ was characterized by
a width already larger than $50$ (in units of $q \sigma_B$ ).

We perform a further analysis considering the variation of the final
$f_q$ with density. The result is shown in Fig.~\ref{fig13}a for a
packing fraction ranging from $0.05$ to $0.50$. The width of the non
ergodicity parameter decreases when the packing fraction is
decreased. This is an indication of the fact that the average
localization length is larger for the lower packing fraction, the one
characterized by a more open structure.  As we discussed above, at
high density the system maintains a certain homogeneity and no open
empty region are detected. An open structure presents large
amplitude modes and, as a consequence, the mean localization length is expected to grow with decreasing packing fraction. This interpretation is confirmed by the mean square displacement (MSD).  In out of equilibrium, this
quantity is defined as $\langle |{\bf r}(t-t_w)-{\bf
r}(t_w)|^2\rangle$, where ${\bf r}(t)$ is the position of the particle
at time $t$ and the average $\langle \cdot
\rangle$ is performed over all the particles. 
Here we focus on the total MSD, i.e. evaluated with no distinction
between particles of the two species for different values of
$t+w$. Results are presented in Fig.~\ref{fig13}b. After a ballistic
short time region, the MSD reaches a plateau, whose value is larger
the smaller $\phi$ is. Since the long-time value of the MSD provides
an estimate of the characteristic size of the cages confining the
particles, data in Fig..~\ref{fig13}b confirm that, in the arrested
state, particles are more and more localized on increasing $\phi$.

\section{Conclusions}

The aim of this manuscript is to provide numerical evidence that, in
the case of SRAC systems, the formation of an arrested state at low packing
fraction results from a phase separation process interrupted by an
attractive glass transition. We have shown that indeed, only when the
system is quenched below $T_x$, the temperature at which the
attractive glass line crosses the coexistence line on the high-polymer
concentration side, the coarsening dynamics get arrested due to the
mobility reduction associated to the glass transition.  We also
observed that the arrest takes place only when the structure is
percolating, which at the studied $T_f$, requires $\phi \gtrsim 0.05$.
For smaller $\phi$, diffusion of the droplets provides a slow
coarsening mechanism which is missing in the percolating case.  For
$T>T_x$, we observed a continuous progressive coarsening process. The
percolating ramified structures produced during the early stages of
the phase separation process become thicker and thicker and the
percolating cluster eventually collapses in the attempt of minimizing
the surface area. These slow rearrangement processes are possible due
to the residual mobility of the high $\phi$ phase.\\
The final structure of the aggregates as a function of $\phi$ is
characterized by a peak at a finite wave-vector $q_p$, which
constitute the frozen memory of the interrupted phase separation
process. The location of the peak and its amplitude depend on
$\phi$. The smaller the $\phi$, the lower is $q_p$ and the larger is
$S(q_p)$.  A complex interaction between the strength of the phase
separation and the mobility in the dense phase define the final
structure of the system. For example, for $\phi=0.50$, the system is
so dense that it freezes before it can get significantly inhomogeneous
and no considerable peak is detected in the static structure
factor. It is also interesting to note that in this case, the
evolution of $S(q)$ is not very different from the case of a quench in
the attractive glass phase region~\cite{Foffi2004b}, i.e. a decrease
in the contact peak with waiting time.  In the low density case the
trend is the opposite, since more particles experience contact after
the quench.
%This some how enforces the idea that attractive glass line (in the MCT fashion) and the gel line are governed by different underlying phenomena~\cite{Foffi2004c}.
It is also worth recalling that the connectivity of the frozen
structure at particle level is also a function of $\phi$.  We find
that the number of bonds is minimized roughly in correspondence of the
critical packing fraction. Perhaps this is related to the strength of
the critical fluctuation, which increase the surface of the aggregate
lowering, as a consequence the number of bonds.\\ Finally we studied
the evolution of the non ergodicity parameter evaluated from the
density-density correlation functions. We found that, differently from
the aging dynamics in glasses, the non ergodicity parameter
progressively increases with $t_w$.  The first components to become
significantly non ergodic are the large wavelength density
fluctuations. The width of $f_q$ increases progressively during the
coarsening dynamics. When dynamic arrest is completed, the
localization length has become extremely narrow. Moreover it changes
with packing fraction, showing less localized aggregates at lower
density.  The progressive increase of the $f_q$ width is reminiscent
of the behavior recently observed in chemical~\cite{Saika2004} and
thermoreversible~\cite{Zaccarelli2004pre} gel formation.  \\ A
final remark is on the effect of the microscopic dynamics. In order to
exclude any artifact introduced by Newtonian dynamics, we performed
simulations, for the lowest $T$, also using Brownian dynamics. The two
microscopic dynamics provide the same equilibrium description but
different time scale. On approaching a structural glass transition, it
has been shown that ND and BD generate the same long time
behavior~\cite{Gleim1998}. When the system is phase separating, as in
our case, the microscopic dynamics could play a major role. We show,
however, that the arrested structure obtained by the two different
dynamical schemes are very similar. Since BD simulation can be a few
order of magnitude slower than ND simulation, it is important to have
the possibility to use the latter to describe aging processes in
colloids.\\

\begin{acknowledgements}
G.~F. acknowledges the support of the Swiss Science Foundation (Grant
No. 99200021-105382/1). Support from MIUR-FIRB and Cofin and Training
Network of the Marie-Curie Programmme of the EU (MRT-CT-2003-504712)
is acknowledged.

\end{acknowledgements}

\bibliography{tesi,add2}

\begin{thebibliography}{63}
\expandafter\ifx\csname natexlab\endcsname\relax\def\natexlab#1{#1}\fi
\expandafter\ifx\csname bibnamefont\endcsname\relax
  \def\bibnamefont#1{#1}\fi
\expandafter\ifx\csname bibfnamefont\endcsname\relax
  \def\bibfnamefont#1{#1}\fi
\expandafter\ifx\csname citenamefont\endcsname\relax
  \def\citenamefont#1{#1}\fi
\expandafter\ifx\csname url\endcsname\relax
  \def\url#1{\texttt{#1}}\fi
\expandafter\ifx\csname urlprefix\endcsname\relax\def\urlprefix{URL }\fi
\providecommand{\bibinfo}[2]{#2}
\providecommand{\eprint}[2][]{\url{#2}}

\bibitem[{\citenamefont{Russel et~al.}(1989)\citenamefont{Russel, Saville, and
  Schowalter}}]{Russel1989}
\bibinfo{author}{\bibfnamefont{W.~B.} \bibnamefont{Russel}},
  \bibinfo{author}{\bibfnamefont{D.~A.} \bibnamefont{Saville}},
  \bibnamefont{and} \bibinfo{author}{\bibfnamefont{W.~R.}
  \bibnamefont{Schowalter}}, \emph{\bibinfo{title}{Colloidal Dispersions}}
  (\bibinfo{publisher}{Cambridge University Press}, \bibinfo{address}{New
  York}, \bibinfo{year}{1989}).

\bibitem[{\citenamefont{Likos}(2001)}]{Likos2001}
\bibinfo{author}{\bibfnamefont{C.~N.} \bibnamefont{Likos}},
  \bibinfo{journal}{Physics Reports} \textbf{\bibinfo{volume}{348}},
  \bibinfo{pages}{267} (\bibinfo{year}{2001}).

\bibitem[{\citenamefont{Anderson and Lekkerkerker}(2002)}]{Anderson2002}
\bibinfo{author}{\bibfnamefont{V.~J.} \bibnamefont{Anderson}} \bibnamefont{and}
  \bibinfo{author}{\bibfnamefont{H.~N.~W.} \bibnamefont{Lekkerkerker}},
  \bibinfo{journal}{Nature} \textbf{\bibinfo{volume}{416}},
  \bibinfo{pages}{811} (\bibinfo{year}{2002}).

\bibitem[{\citenamefont{Cipelletti}(2005)}]{Cipelletti2005}
\bibinfo{author}{\bibfnamefont{L.}~\bibnamefont{Cipelletti}}
  (\bibinfo{year}{2005}), \bibinfo{note}{to be published in J.~Phys.:
  Condens.~Matter}.

\bibitem[{\citenamefont{Frenkel}(2002)}]{Frenkel2002}
\bibinfo{author}{\bibfnamefont{D.}~\bibnamefont{Frenkel}},
  \bibinfo{journal}{Science} \textbf{\bibinfo{volume}{296}},
  \bibinfo{pages}{65} (\bibinfo{year}{2002}).

\bibitem[{\citenamefont{Sciortino}(2002)}]{Sciortino2002}
\bibinfo{author}{\bibfnamefont{F.}~\bibnamefont{Sciortino}},
  \bibinfo{journal}{Nature Materials} \textbf{\bibinfo{volume}{1}},
  \bibinfo{pages}{145} (\bibinfo{year}{2002}).

\bibitem[{\citenamefont{A.P.Gast et~al.}(1983)\citenamefont{A.P.Gast, Russell,
  and Hall}}]{Gast83}
\bibinfo{author}{\bibnamefont{A.P.Gast}},
  \bibinfo{author}{\bibfnamefont{W.}~\bibnamefont{Russell}}, \bibnamefont{and}
  \bibinfo{author}{\bibfnamefont{C.}~\bibnamefont{Hall}}, \bibinfo{journal}{J.
  Colloid Interface Sci.} \textbf{\bibinfo{volume}{96}}, \bibinfo{pages}{1977}
  (\bibinfo{year}{1983}).

\bibitem[{\citenamefont{Tejero et~al.}(1994)\citenamefont{Tejero, Daanoun,
  Lekkerkerker, and Baus}}]{Tejero1994}
\bibinfo{author}{\bibfnamefont{C.~F.} \bibnamefont{Tejero}},
  \bibinfo{author}{\bibfnamefont{A.}~\bibnamefont{Daanoun}},
  \bibinfo{author}{\bibfnamefont{H.~N.~W.} \bibnamefont{Lekkerkerker}},
  \bibnamefont{and} \bibinfo{author}{\bibfnamefont{M.}~\bibnamefont{Baus}},
  \bibinfo{journal}{Phys.~Rev.~Lett.} \textbf{\bibinfo{volume}{73}},
  \bibinfo{pages}{752} (\bibinfo{year}{1994}).

\bibitem[{\citenamefont{Dijkstra et~al.}(1999)\citenamefont{Dijkstra, Brader,
  and Evans}}]{Dijkstra1999}
\bibinfo{author}{\bibfnamefont{M.}~\bibnamefont{Dijkstra}},
  \bibinfo{author}{\bibfnamefont{J.}~\bibnamefont{Brader}}, \bibnamefont{and}
  \bibinfo{author}{\bibfnamefont{R.}~\bibnamefont{Evans}},
  \bibinfo{journal}{J.Phys.:Condens. Matter} \textbf{\bibinfo{volume}{11}},
  \bibinfo{pages}{10079} (\bibinfo{year}{1999}).

\bibitem[{\citenamefont{Hagen and Frenkel}(1994)}]{Hagen1994}
\bibinfo{author}{\bibfnamefont{M.}~\bibnamefont{Hagen}} \bibnamefont{and}
  \bibinfo{author}{\bibfnamefont{D.}~\bibnamefont{Frenkel}},
  \bibinfo{journal}{J. Chem.Phys} \textbf{\bibinfo{volume}{101}},
  \bibinfo{pages}{4093} (\bibinfo{year}{1994}).

\bibitem[{\citenamefont{Lekkerkerker et~al.}(1992)\citenamefont{Lekkerkerker,
  Poon, Pusey, Stroobants, and Warren}}]{Lekkerkerker1992}
\bibinfo{author}{\bibfnamefont{H.~N.~W.} \bibnamefont{Lekkerkerker}},
  \bibinfo{author}{\bibfnamefont{W.~C.~K.} \bibnamefont{Poon}},
  \bibinfo{author}{\bibfnamefont{P.~N.} \bibnamefont{Pusey}},
  \bibinfo{author}{\bibfnamefont{A.}~\bibnamefont{Stroobants}},
  \bibnamefont{and} \bibinfo{author}{\bibfnamefont{P.~B.}
  \bibnamefont{Warren}}, \bibinfo{journal}{Europhys.~Lett.}
  \textbf{\bibinfo{volume}{20}}, \bibinfo{pages}{559} (\bibinfo{year}{1992}).

\bibitem[{\citenamefont{Illett et~al.}(1995)\citenamefont{Illett, Orrock, Poon,
  and P.N.Pusey}}]{Illet1995}
\bibinfo{author}{\bibfnamefont{S.}~\bibnamefont{Illett}},
  \bibinfo{author}{\bibfnamefont{A.}~\bibnamefont{Orrock}},
  \bibinfo{author}{\bibfnamefont{W.}~\bibnamefont{Poon}}, \bibnamefont{and}
  \bibinfo{author}{\bibnamefont{P.N.Pusey}}, \bibinfo{journal}{Phys. Rev E}
  \textbf{\bibinfo{volume}{51}}, \bibinfo{pages}{1344} (\bibinfo{year}{1995}).

\bibitem[{\citenamefont{Auer and Frenkel}(2001)}]{Auer2001}
\bibinfo{author}{\bibfnamefont{S.}~\bibnamefont{Auer}} \bibnamefont{and}
  \bibinfo{author}{\bibfnamefont{D.}~\bibnamefont{Frenkel}},
  \bibinfo{journal}{Nature} \textbf{\bibinfo{volume}{409}},
  \bibinfo{pages}{1020} (\bibinfo{year}{2001}).

\bibitem[{\citenamefont{Foffi et~al.}(2002)\citenamefont{Foffi, Dawson,
  Buldyrev, Sciortino, Zaccarelli, and Tartaglia}}]{Foffi2002c}
\bibinfo{author}{\bibfnamefont{G.}~\bibnamefont{Foffi}},
  \bibinfo{author}{\bibfnamefont{K.~A.} \bibnamefont{Dawson}},
  \bibinfo{author}{\bibfnamefont{S.~V.} \bibnamefont{Buldyrev}},
  \bibinfo{author}{\bibfnamefont{F.}~\bibnamefont{Sciortino}},
  \bibinfo{author}{\bibfnamefont{E.}~\bibnamefont{Zaccarelli}},
  \bibnamefont{and}
  \bibinfo{author}{\bibfnamefont{P.}~\bibnamefont{Tartaglia}},
  \bibinfo{journal}{Phys.~Rev.~E} \textbf{\bibinfo{volume}{65}},
  \bibinfo{pages}{050802(R)} (\bibinfo{year}{2002}).

\bibitem[{\citenamefont{Zaccarelli et~al.}(2002)\citenamefont{Zaccarelli,
  Foffi, Dawson, Buldyrev, Sciortino, and Tartaglia}}]{Zaccarelli2002b}
\bibinfo{author}{\bibfnamefont{E.}~\bibnamefont{Zaccarelli}},
  \bibinfo{author}{\bibfnamefont{G.}~\bibnamefont{Foffi}},
  \bibinfo{author}{\bibfnamefont{K.~A.} \bibnamefont{Dawson}},
  \bibinfo{author}{\bibfnamefont{S.~V.} \bibnamefont{Buldyrev}},
  \bibinfo{author}{\bibfnamefont{F.}~\bibnamefont{Sciortino}},
  \bibnamefont{and}
  \bibinfo{author}{\bibfnamefont{P.}~\bibnamefont{Tartaglia}},
  \bibinfo{journal}{Phys.~Rev.~E} \textbf{\bibinfo{volume}{66}},
  \bibinfo{pages}{041402} (\bibinfo{year}{2002}).

\bibitem[{\citenamefont{Fabbian et~al.}(1999)\citenamefont{Fabbian, G{\"o}tze,
  Sciortino, Tartaglia, and Thiery}}]{Fabbian1999}
\bibinfo{author}{\bibfnamefont{L.}~\bibnamefont{Fabbian}},
  \bibinfo{author}{\bibfnamefont{W.}~\bibnamefont{G{\"o}tze}},
  \bibinfo{author}{\bibfnamefont{F.}~\bibnamefont{Sciortino}},
  \bibinfo{author}{\bibfnamefont{P.}~\bibnamefont{Tartaglia}},
  \bibnamefont{and} \bibinfo{author}{\bibfnamefont{F.}~\bibnamefont{Thiery}},
  \bibinfo{journal}{Phys.~Rev.~E} \textbf{\bibinfo{volume}{59}},
  \bibinfo{pages}{R1347} (\bibinfo{year}{1999}).

\bibitem[{\citenamefont{Bergenholtz and Fuchs}(1999)}]{Bergenholtz1999}
\bibinfo{author}{\bibfnamefont{J.}~\bibnamefont{Bergenholtz}} \bibnamefont{and}
  \bibinfo{author}{\bibfnamefont{M.}~\bibnamefont{Fuchs}},
  \bibinfo{journal}{Phys.~Rev.~E} \textbf{\bibinfo{volume}{59}},
  \bibinfo{pages}{5706} (\bibinfo{year}{1999}).

\bibitem[{\citenamefont{Dawson et~al.}(2001)\citenamefont{Dawson, Foffi, Fuchs,
  G\"otze, Sciortino, Sperl, Tartaglia, Voigtmann, and
  Zaccarelli}}]{Dawson2001}
\bibinfo{author}{\bibfnamefont{K.~A.} \bibnamefont{Dawson}},
  \bibinfo{author}{\bibfnamefont{G.}~\bibnamefont{Foffi}},
  \bibinfo{author}{\bibfnamefont{M.}~\bibnamefont{Fuchs}},
  \bibinfo{author}{\bibfnamefont{W.}~\bibnamefont{G\"otze}},
  \bibinfo{author}{\bibfnamefont{F.}~\bibnamefont{Sciortino}},
  \bibinfo{author}{\bibfnamefont{M.}~\bibnamefont{Sperl}},
  \bibinfo{author}{\bibfnamefont{P.}~\bibnamefont{Tartaglia}},
  \bibinfo{author}{\bibfnamefont{T.}~\bibnamefont{Voigtmann}},
  \bibnamefont{and}
  \bibinfo{author}{\bibfnamefont{E.}~\bibnamefont{Zaccarelli}},
  \bibinfo{journal}{Phys.~Rev.~E} \textbf{\bibinfo{volume}{63}},
  \bibinfo{pages}{011401} (\bibinfo{year}{2001}).

\bibitem[{\citenamefont{Sperl}(2003{\natexlab{a}})}]{Sperl2003b}
\bibinfo{author}{\bibfnamefont{M.}~\bibnamefont{Sperl}},
  \bibinfo{journal}{Phys. Rev. E} \textbf{\bibinfo{volume}{68}},
  \bibinfo{pages}{031405} (\bibinfo{year}{2003}{\natexlab{a}}).

\bibitem[{\citenamefont{Foffi et~al.}(2004{\natexlab{a}})\citenamefont{Foffi,
  Zaccarelli, Buldyrev, Sciortino, and Tartaglia}}]{Foffi2004b}
\bibinfo{author}{\bibfnamefont{G.}~\bibnamefont{Foffi}},
  \bibinfo{author}{\bibfnamefont{E.}~\bibnamefont{Zaccarelli}},
  \bibinfo{author}{\bibfnamefont{S.}~\bibnamefont{Buldyrev}},
  \bibinfo{author}{\bibfnamefont{F.}~\bibnamefont{Sciortino}},
  \bibnamefont{and}
  \bibinfo{author}{\bibfnamefont{P.}~\bibnamefont{Tartaglia}},
  \bibinfo{journal}{J. Chem. Phys.} \textbf{\bibinfo{volume}{120}},
  \bibinfo{pages}{8824} (\bibinfo{year}{2004}{\natexlab{a}}).

\bibitem[{\citenamefont{G\"otze and Sperl}(2004)}]{Gotze2004}
\bibinfo{author}{\bibfnamefont{W.}~\bibnamefont{G\"otze}} \bibnamefont{and}
  \bibinfo{author}{\bibfnamefont{M.}~\bibnamefont{Sperl}},
  \bibinfo{journal}{J.~Phys.: Condens.~Matter} \textbf{\bibinfo{volume}{16}},
  \bibinfo{pages}{S4807} (\bibinfo{year}{2004}).

\bibitem[{\citenamefont{G{\"o}tze}(1991)}]{Goetze1991b}
\bibinfo{author}{\bibfnamefont{W.}~\bibnamefont{G{\"o}tze}}, in
  \emph{\bibinfo{booktitle}{Liquids, Freezing and Glass Transition}}, edited by
  \bibinfo{editor}{\bibfnamefont{J.~P.} \bibnamefont{Hansen}},
  \bibinfo{editor}{\bibfnamefont{D.}~\bibnamefont{Levesque}}, \bibnamefont{and}
  \bibinfo{editor}{\bibfnamefont{J.}~\bibnamefont{Zinn-Justin}}
  (\bibinfo{publisher}{North Holland}, \bibinfo{address}{Amsterdam},
  \bibinfo{year}{1991}), vol. \bibinfo{volume}{Session LI (1989)} of
  \emph{\bibinfo{series}{Les Houches Summer Schools of Theoretical Physics}},
  pp. \bibinfo{pages}{287--503}.

\bibitem[{\citenamefont{Puertas et~al.}(2002)\citenamefont{Puertas, Fuchs, and
  Cates}}]{Puertas2002}
\bibinfo{author}{\bibfnamefont{A.~M.} \bibnamefont{Puertas}},
  \bibinfo{author}{\bibfnamefont{M.}~\bibnamefont{Fuchs}}, \bibnamefont{and}
  \bibinfo{author}{\bibfnamefont{M.~E.} \bibnamefont{Cates}},
  \bibinfo{journal}{Phys.~Rev.~Lett.} \textbf{\bibinfo{volume}{88}},
  \bibinfo{pages}{098301} (\bibinfo{year}{2002}).

\bibitem[{\citenamefont{Sciortino et~al.}(2003)\citenamefont{Sciortino,
  Tartaglia, and Zaccarelli}}]{Sciortino2003}
\bibinfo{author}{\bibfnamefont{F.}~\bibnamefont{Sciortino}},
  \bibinfo{author}{\bibfnamefont{P.}~\bibnamefont{Tartaglia}},
  \bibnamefont{and}
  \bibinfo{author}{\bibfnamefont{E.}~\bibnamefont{Zaccarelli}},
  \bibinfo{journal}{Phys. Rev. Lett.} \textbf{\bibinfo{volume}{91}},
  \bibinfo{pages}{268301} (\bibinfo{year}{2003}).

\bibitem[{\citenamefont{Mallamace et~al.}(2000)\citenamefont{Mallamace,
  Gambadauro, Micali, Tartaglia, Liao, and Chen}}]{Mallamace2000}
\bibinfo{author}{\bibfnamefont{F.}~\bibnamefont{Mallamace}},
  \bibinfo{author}{\bibfnamefont{P.}~\bibnamefont{Gambadauro}},
  \bibinfo{author}{\bibfnamefont{N.}~\bibnamefont{Micali}},
  \bibinfo{author}{\bibfnamefont{P.}~\bibnamefont{Tartaglia}},
  \bibinfo{author}{\bibfnamefont{C.}~\bibnamefont{Liao}}, \bibnamefont{and}
  \bibinfo{author}{\bibfnamefont{S.-H.} \bibnamefont{Chen}},
  \bibinfo{journal}{Phys.~Rev.~Lett.} \textbf{\bibinfo{volume}{84}},
  \bibinfo{pages}{5431} (\bibinfo{year}{2000}).

\bibitem[{\citenamefont{Pham et~al.}(2002)\citenamefont{Pham, Puertas,
  Bergenholtz, Egelhaaf, Moussa\"{\i}d, Pusey, Schofield, Cates, Fuchs, and
  Poon}}]{Pham2002}
\bibinfo{author}{\bibfnamefont{K.~N.} \bibnamefont{Pham}},
  \bibinfo{author}{\bibfnamefont{A.~M.} \bibnamefont{Puertas}},
  \bibinfo{author}{\bibfnamefont{J.}~\bibnamefont{Bergenholtz}},
  \bibinfo{author}{\bibfnamefont{S.~U.} \bibnamefont{Egelhaaf}},
  \bibinfo{author}{\bibfnamefont{A.}~\bibnamefont{Moussa\"{\i}d}},
  \bibinfo{author}{\bibfnamefont{P.~N.} \bibnamefont{Pusey}},
  \bibinfo{author}{\bibfnamefont{A.~B.} \bibnamefont{Schofield}},
  \bibinfo{author}{\bibfnamefont{M.~E.} \bibnamefont{Cates}},
  \bibinfo{author}{\bibfnamefont{M.}~\bibnamefont{Fuchs}}, \bibnamefont{and}
  \bibinfo{author}{\bibfnamefont{W.~C.~K.} \bibnamefont{Poon}},
  \bibinfo{journal}{Science} \textbf{\bibinfo{volume}{296}},
  \bibinfo{pages}{104} (\bibinfo{year}{2002}).

\bibitem[{\citenamefont{Eckert and Bartsch}(2002)}]{Eckert2002}
\bibinfo{author}{\bibfnamefont{T.}~\bibnamefont{Eckert}} \bibnamefont{and}
  \bibinfo{author}{\bibfnamefont{E.}~\bibnamefont{Bartsch}},
  \bibinfo{journal}{Phys.~Rev.~Lett.} \textbf{\bibinfo{volume}{89}},
  \bibinfo{pages}{125701} (\bibinfo{year}{2002}).

\bibitem[{\citenamefont{Chen et~al.}(2003)\citenamefont{Chen, Chen, and
  Mallamace}}]{Chen2003}
\bibinfo{author}{\bibfnamefont{S.-H.} \bibnamefont{Chen}},
  \bibinfo{author}{\bibfnamefont{W.-R.} \bibnamefont{Chen}}, \bibnamefont{and}
  \bibinfo{author}{\bibfnamefont{F.}~\bibnamefont{Mallamace}},
  \bibinfo{journal}{Science} \textbf{\bibinfo{volume}{300}},
  \bibinfo{pages}{619} (\bibinfo{year}{2003}).

\bibitem[{\citenamefont{Poon}(1998)}]{Poon1998}
\bibinfo{author}{\bibfnamefont{W.~C.~K.} \bibnamefont{Poon}},
  \bibinfo{journal}{Curr.~Opin.~Colloid Interf.~Sci.}
  \textbf{\bibinfo{volume}{3}}, \bibinfo{pages}{593} (\bibinfo{year}{1998}).

\bibitem[{\citenamefont{Verhaeg et~al.}(1999)\citenamefont{Verhaeg, Asnaghi,
  and Lekkerkerker}}]{Verhaegh1999}
\bibinfo{author}{\bibfnamefont{N.~A.~M.} \bibnamefont{Verhaeg}},
  \bibinfo{author}{\bibfnamefont{D.}~\bibnamefont{Asnaghi}}, \bibnamefont{and}
  \bibinfo{author}{\bibfnamefont{H.}~\bibnamefont{Lekkerkerker}},
  \bibinfo{journal}{Physica A} \textbf{\bibinfo{volume}{264}},
  \bibinfo{pages}{64} (\bibinfo{year}{1999}).

\bibitem[{\citenamefont{Segr{\`e} et~al.}(2001)\citenamefont{Segr{\`e}, Prasad,
  Schofield, and Weitz}}]{Segre2001}
\bibinfo{author}{\bibfnamefont{P.~N.} \bibnamefont{Segr{\`e}}},
  \bibinfo{author}{\bibfnamefont{V.}~\bibnamefont{Prasad}},
  \bibinfo{author}{\bibfnamefont{A.~B.} \bibnamefont{Schofield}},
  \bibnamefont{and} \bibinfo{author}{\bibfnamefont{D.~A.} \bibnamefont{Weitz}},
  \bibinfo{journal}{Phys.~Rev.~Lett.} \textbf{\bibinfo{volume}{86}},
  \bibinfo{pages}{6042} (\bibinfo{year}{2001}).

\bibitem[{\citenamefont{Shah et~al.}(2003)\citenamefont{Shah, Chen,
  Ramakrishnan, Schweizer, and Zukoski}}]{Shah2003}
\bibinfo{author}{\bibfnamefont{S.~A.} \bibnamefont{Shah}},
  \bibinfo{author}{\bibfnamefont{Y.-L.} \bibnamefont{Chen}},
  \bibinfo{author}{\bibfnamefont{S.}~\bibnamefont{Ramakrishnan}},
  \bibinfo{author}{\bibfnamefont{K.~S.} \bibnamefont{Schweizer}},
  \bibnamefont{and} \bibinfo{author}{\bibfnamefont{C.~F.}
  \bibnamefont{Zukoski}}, \bibinfo{journal}{J.~Phys.: Condens. Matter}
  \textbf{\bibinfo{volume}{15}}, \bibinfo{pages}{4751} (\bibinfo{year}{2003}).

\bibitem[{\citenamefont{Sedgwick
  et~al.}(2004{\natexlab{a}})\citenamefont{Sedgwick, Egelhaaf, and
  Poon}}]{Sedgqick2004}
\bibinfo{author}{\bibfnamefont{H.}~\bibnamefont{Sedgwick}},
  \bibinfo{author}{\bibfnamefont{S.~U.} \bibnamefont{Egelhaaf}},
  \bibnamefont{and} \bibinfo{author}{\bibfnamefont{W.~C.~K.}
  \bibnamefont{Poon}}, \bibinfo{journal}{J.~Phys.: Condens.~Matter}
  \textbf{\bibinfo{volume}{16}}, \bibinfo{pages}{S4913}
  (\bibinfo{year}{2004}{\natexlab{a}}).

\bibitem[{\citenamefont{Trappe and Sandk{\"u}hler}(2004)}]{Trappe2004}
\bibinfo{author}{\bibfnamefont{V.}~\bibnamefont{Trappe}} \bibnamefont{and}
  \bibinfo{author}{\bibfnamefont{P.}~\bibnamefont{Sandk{\"u}hler}},
  \bibinfo{journal}{Curr. Opin. Colloid Int. Sci.}
  \textbf{\bibinfo{volume}{8}}, \bibinfo{pages}{494} (\bibinfo{year}{2004}).

\bibitem[{\citenamefont{Sciortino et~al.}(2004)\citenamefont{Sciortino, Mossa,
  Zaccarelli, and Tartaglia}}]{Sciortino2004}
\bibinfo{author}{\bibfnamefont{F.}~\bibnamefont{Sciortino}},
  \bibinfo{author}{\bibfnamefont{S.}~\bibnamefont{Mossa}},
  \bibinfo{author}{\bibfnamefont{E.}~\bibnamefont{Zaccarelli}},
  \bibnamefont{and}
  \bibinfo{author}{\bibfnamefont{P.}~\bibnamefont{Tartaglia}},
  \bibinfo{journal}{Phys.~Rev.~Lett.} \textbf{\bibinfo{volume}{93}},
  \bibinfo{pages}{055701} (\bibinfo{year}{2004}).

\bibitem[{\citenamefont{Sator et~al.}(2004)\citenamefont{Sator, Fierro, Gado,
  and Coniglio}}]{coniglio}
\bibinfo{author}{\bibfnamefont{N.}~\bibnamefont{Sator}},
  \bibinfo{author}{\bibfnamefont{A.}~\bibnamefont{Fierro}},
  \bibinfo{author}{\bibfnamefont{E.~D.} \bibnamefont{Gado}}, \bibnamefont{and}
  \bibinfo{author}{\bibfnamefont{A.}~\bibnamefont{Coniglio}}
  (\bibinfo{year}{2004}), \eprint{cond-mat/0309007}.

\bibitem[{\citenamefont{Kroy et~al.}(2004)\citenamefont{Kroy, Cates, and
  Poon}}]{Kroy2004}
\bibinfo{author}{\bibfnamefont{K.}~\bibnamefont{Kroy}},
  \bibinfo{author}{\bibfnamefont{M.~E.} \bibnamefont{Cates}}, \bibnamefont{and}
  \bibinfo{author}{\bibfnamefont{W.~C.~K.} \bibnamefont{Poon}},
  \bibinfo{journal}{Phys.~Rev.~Lett.} \textbf{\bibinfo{volume}{92}},
  \bibinfo{pages}{148302} (\bibinfo{year}{2004}).

\bibitem[{\citenamefont{Bos and van Opheusden}(1996)}]{Bos1996}
\bibinfo{author}{\bibfnamefont{M.~T.~A.} \bibnamefont{Bos}} \bibnamefont{and}
  \bibinfo{author}{\bibfnamefont{J.~H.~J.} \bibnamefont{van Opheusden}},
  \bibinfo{journal}{Phys.~Rev. E} \textbf{\bibinfo{volume}{53}},
  \bibinfo{pages}{5044} (\bibinfo{year}{1996}).

\bibitem[{\citenamefont{Soga et~al.}(1998)\citenamefont{Soga, Melrose, and
  Ball}}]{Soga1998}
\bibinfo{author}{\bibfnamefont{K.~G.} \bibnamefont{Soga}},
  \bibinfo{author}{\bibfnamefont{J.~R.} \bibnamefont{Melrose}},
  \bibnamefont{and} \bibinfo{author}{\bibfnamefont{R.~C.} \bibnamefont{Ball}},
  \bibinfo{journal}{J Chem. Phys.} \textbf{\bibinfo{volume}{108}},
  \bibinfo{pages}{6026} (\bibinfo{year}{1998}).

\bibitem[{\citenamefont{Lodge and Heyes}(1999)}]{Lodge1999}
\bibinfo{author}{\bibfnamefont{J.}~\bibnamefont{Lodge}} \bibnamefont{and}
  \bibinfo{author}{\bibfnamefont{D.}~\bibnamefont{Heyes}},
  \bibinfo{journal}{Phys. Chem. Chem. Phys} \textbf{\bibinfo{volume}{1}},
  \bibinfo{pages}{2119} (\bibinfo{year}{1999}).

\bibitem[{\citenamefont{d'Arjuzon et~al.}(2003)\citenamefont{d'Arjuzon, Frith,
  and Melrose}}]{Arjuzon2003}
\bibinfo{author}{\bibfnamefont{R.~J.~M.} \bibnamefont{d'Arjuzon}},
  \bibinfo{author}{\bibfnamefont{W.}~\bibnamefont{Frith}}, \bibnamefont{and}
  \bibinfo{author}{\bibfnamefont{J.~R.} \bibnamefont{Melrose}},
  \bibinfo{journal}{Phys.~Rev. E} \textbf{\bibinfo{volume}{67}},
  \bibinfo{pages}{061404} (\bibinfo{year}{2003}).

\bibitem[{\citenamefont{Zaccarelli et~al.}(2004)\citenamefont{Zaccarelli,
  Sciortino, Buldyrev, and Tartaglia}}]{Zaccarelli2004}
\bibinfo{author}{\bibfnamefont{E.}~\bibnamefont{Zaccarelli}},
  \bibinfo{author}{\bibfnamefont{F.}~\bibnamefont{Sciortino}},
  \bibinfo{author}{\bibfnamefont{S.}~\bibnamefont{Buldyrev}}, \bibnamefont{and}
  \bibinfo{author}{\bibfnamefont{P.}~\bibnamefont{Tartaglia}}, in
  \emph{\bibinfo{booktitle}{Unifyng Concepts in Granular Media and Glasses}},
  edited by \bibinfo{editor}{\bibfnamefont{A.}~\bibnamefont{Coniglio}},
  \bibinfo{editor}{\bibfnamefont{A.}~\bibnamefont{Fierro}},
  \bibinfo{editor}{\bibfnamefont{H.}~\bibnamefont{Herrmann}}, \bibnamefont{and}
  \bibinfo{editor}{\bibfnamefont{M.}~\bibnamefont{Nicodemi}}
  (\bibinfo{publisher}{Elsevier}, \bibinfo{address}{Amsterdam},
  \bibinfo{year}{2004}).

\bibitem[{\citenamefont{Foffi et~al.}()\citenamefont{Foffi, \surname{De
  Michele}, F.Sciortino, and Tartaglia}}]{Foffi2004cpre}
\bibinfo{author}{\bibfnamefont{G.}~\bibnamefont{Foffi}},
  \bibinfo{author}{\bibfnamefont{C.}~\bibnamefont{\surname{De Michele}}},
  \bibinfo{author}{\bibnamefont{F.Sciortino}}, \bibnamefont{and}
  \bibinfo{author}{\bibfnamefont{P.}~\bibnamefont{Tartaglia}},
  \bibinfo{note}{cond-mat/0410358}.

\bibitem[{\citenamefont{Manley et~al.}(2005)\citenamefont{Manley, Wyss,
  Miyazaki, Conrad, Trappe, Kaufman, Reichman, and D.A.Weitz}}]{Manley2005}
\bibinfo{author}{\bibfnamefont{S.}~\bibnamefont{Manley}},
  \bibinfo{author}{\bibfnamefont{H.~M.} \bibnamefont{Wyss}},
  \bibinfo{author}{\bibfnamefont{K.}~\bibnamefont{Miyazaki}},
  \bibinfo{author}{\bibfnamefont{J.}~\bibnamefont{Conrad}},
  \bibinfo{author}{\bibfnamefont{V.}~\bibnamefont{Trappe}},
  \bibinfo{author}{\bibfnamefont{L.}~\bibnamefont{Kaufman}},
  \bibinfo{author}{\bibfnamefont{D.}~\bibnamefont{Reichman}}, \bibnamefont{and}
  \bibinfo{author}{\bibnamefont{D.A.Weitz}} (\bibinfo{year}{2005}),
  \eprint{cond-mat/0309007}.

\bibitem[{\citenamefont{Noro and Frenkel}(2000)}]{Noro2000}
\bibinfo{author}{\bibfnamefont{M.}~\bibnamefont{Noro}} \bibnamefont{and}
  \bibinfo{author}{\bibfnamefont{D.}~\bibnamefont{Frenkel}},
  \bibinfo{journal}{J.Chem.Phys.} \textbf{\bibinfo{volume}{113}},
  \bibinfo{pages}{2941} (\bibinfo{year}{2000}).

\bibitem[{\citenamefont{Rapaport}(1997)}]{Rapaport97}
\bibinfo{author}{\bibfnamefont{D.~C.} \bibnamefont{Rapaport}},
  \emph{\bibinfo{title}{The art of computer simulations}}
  (\bibinfo{publisher}{Cambridge Univ Press}, \bibinfo{address}{London},
  \bibinfo{year}{1997}), \bibinfo{edition}{2nd} ed.

\bibitem[{\citenamefont{Strating}(1999)}]{Strating1999}
\bibinfo{author}{\bibfnamefont{P.}~\bibnamefont{Strating}},
  \bibinfo{journal}{Phys.~Rev.~E} \textbf{\bibinfo{volume}{59}},
  \bibinfo{pages}{2175} (\bibinfo{year}{1999}).

\bibitem[{\citenamefont{Baxter}(1968)}]{Baxter1968b}
\bibinfo{author}{\bibfnamefont{R.~J.} \bibnamefont{Baxter}},
  \bibinfo{journal}{J.~Chem.~Phys.} \textbf{\bibinfo{volume}{49}},
  \bibinfo{pages}{2770} (\bibinfo{year}{1968}).

\bibitem[{\citenamefont{Miller and Frenkel}(2003)}]{Miller2003}
\bibinfo{author}{\bibfnamefont{M.~A.} \bibnamefont{Miller}} \bibnamefont{and}
  \bibinfo{author}{\bibfnamefont{D.}~\bibnamefont{Frenkel}},
  \bibinfo{journal}{Phys.~Rev.~Lett.} \textbf{\bibinfo{volume}{90}},
  \bibinfo{pages}{135702} (\bibinfo{year}{2003}).

\bibitem[{\citenamefont{Glotzer et~al.}(1994)\citenamefont{Glotzer, Gyure,
  Sciortino, Coniglio, and Stanley}}]{Glotzer1994}
\bibinfo{author}{\bibfnamefont{S.~C.} \bibnamefont{Glotzer}},
  \bibinfo{author}{\bibfnamefont{M.~F.} \bibnamefont{Gyure}},
  \bibinfo{author}{\bibfnamefont{F.}~\bibnamefont{Sciortino}},
  \bibinfo{author}{\bibfnamefont{A.}~\bibnamefont{Coniglio}}, \bibnamefont{and}
  \bibinfo{author}{\bibfnamefont{H.~E.} \bibnamefont{Stanley}},
  \bibinfo{journal}{Phys.~Rev.~E} \textbf{\bibinfo{volume}{49}},
  \bibinfo{pages}{247} (\bibinfo{year}{1994}).

\bibitem[{\citenamefont{Zaccarelli et~al.}(2003)\citenamefont{Zaccarelli,
  Foffi, Dawson, Buldyrev, Sciortino, and Tartaglia}}]{Zaccarelli2003}
\bibinfo{author}{\bibfnamefont{E.}~\bibnamefont{Zaccarelli}},
  \bibinfo{author}{\bibfnamefont{G.}~\bibnamefont{Foffi}},
  \bibinfo{author}{\bibfnamefont{K.~A.} \bibnamefont{Dawson}},
  \bibinfo{author}{\bibfnamefont{S.~V.} \bibnamefont{Buldyrev}},
  \bibinfo{author}{\bibfnamefont{F.}~\bibnamefont{Sciortino}},
  \bibnamefont{and}
  \bibinfo{author}{\bibfnamefont{P.}~\bibnamefont{Tartaglia}},
  \bibinfo{journal}{J.~Phys.: Condens.~Matter} \textbf{\bibinfo{volume}{16}},
  \bibinfo{pages}{S367} (\bibinfo{year}{2003}).

\bibitem[{\citenamefont{Weeks et~al.}(2000)\citenamefont{Weeks, Crocker,
  Levitt, Schofield, and Weitz}}]{Weeks2000}
\bibinfo{author}{\bibfnamefont{E.~R.} \bibnamefont{Weeks}},
  \bibinfo{author}{\bibfnamefont{J.~C.} \bibnamefont{Crocker}},
  \bibinfo{author}{\bibfnamefont{A.~C.} \bibnamefont{Levitt}},
  \bibinfo{author}{\bibfnamefont{A.}~\bibnamefont{Schofield}},
  \bibnamefont{and} \bibinfo{author}{\bibfnamefont{D.~A.} \bibnamefont{Weitz}},
  \bibinfo{journal}{Science} \textbf{\bibinfo{volume}{287}},
  \bibinfo{pages}{627} (\bibinfo{year}{2000}).

\bibitem[{\citenamefont{U.Gasser et~al.}(2001)\citenamefont{U.Gasser, Weeks,
  Schofield, Pussey, and Weitz}}]{Gasser2001}
\bibinfo{author}{\bibnamefont{U.Gasser}},
  \bibinfo{author}{\bibfnamefont{E.}~\bibnamefont{Weeks}},
  \bibinfo{author}{\bibfnamefont{A.}~\bibnamefont{Schofield}},
  \bibinfo{author}{\bibfnamefont{P.}~\bibnamefont{Pussey}}, \bibnamefont{and}
  \bibinfo{author}{\bibfnamefont{D.}~\bibnamefont{Weitz}},
  \bibinfo{journal}{Science} \textbf{\bibinfo{volume}{292}},
  \bibinfo{pages}{258} (\bibinfo{year}{2001}).

\bibitem[{\citenamefont{Aarts et~al.}(2004)\citenamefont{Aarts, Schmidt, and
  Lekkerkerker}}]{Aarts2004}
\bibinfo{author}{\bibfnamefont{D.~G. A.~L.} \bibnamefont{Aarts}},
  \bibinfo{author}{\bibfnamefont{M.}~\bibnamefont{Schmidt}}, \bibnamefont{and}
  \bibinfo{author}{\bibfnamefont{H.~N.~W.} \bibnamefont{Lekkerkerker}},
  \bibinfo{journal}{Science} \textbf{\bibinfo{volume}{304}},
  \bibinfo{pages}{847} (\bibinfo{year}{2004}).

\bibitem[{\citenamefont{Kilfoil et~al.}(2003)\citenamefont{Kilfoil, Pashkovski,
  Masters, and Weitz}}]{Kilfoil2003}
\bibinfo{author}{\bibfnamefont{M.~L.} \bibnamefont{Kilfoil}},
  \bibinfo{author}{\bibfnamefont{E.~E.} \bibnamefont{Pashkovski}},
  \bibinfo{author}{\bibfnamefont{J.~A.} \bibnamefont{Masters}},
  \bibnamefont{and} \bibinfo{author}{\bibfnamefont{D.~A.} \bibnamefont{Weitz}},
  \bibinfo{journal}{Philosophical Transactions: Mathematical, Physical and
  Engineering Sciences} \textbf{\bibinfo{volume}{361}}, \bibinfo{pages}{753}
  (\bibinfo{year}{2003}).

\bibitem[{\citenamefont{Sedgwick
  et~al.}(2004{\natexlab{b}})\citenamefont{Sedgwick, Egelhaaf, , and
  Poon}}]{Poon2004}
\bibinfo{author}{\bibfnamefont{H.}~\bibnamefont{Sedgwick}},
  \bibinfo{author}{\bibfnamefont{S.~U.} \bibnamefont{Egelhaaf}}, ,
  \bibnamefont{and} \bibinfo{author}{\bibfnamefont{W.~C.~K.}
  \bibnamefont{Poon}}, \bibinfo{journal}{J. Phys: Cond. Mat.}
  \textbf{\bibinfo{volume}{16}}, \bibinfo{pages}{S4913}
  (\bibinfo{year}{2004}{\natexlab{b}}).

\bibitem[{\citenamefont{Stradner et~al.}(2004)\citenamefont{Stradner, Sedgwick,
  Cardinaux, Poon, Egelhaaf, and Schurtenberger}}]{Stradner2004}
\bibinfo{author}{\bibfnamefont{A.}~\bibnamefont{Stradner}},
  \bibinfo{author}{\bibfnamefont{H.}~\bibnamefont{Sedgwick}},
  \bibinfo{author}{\bibfnamefont{F.}~\bibnamefont{Cardinaux}},
  \bibinfo{author}{\bibfnamefont{W.~C.~K.} \bibnamefont{Poon}},
  \bibinfo{author}{\bibfnamefont{S.~U.} \bibnamefont{Egelhaaf}},
  \bibnamefont{and}
  \bibinfo{author}{\bibfnamefont{P.}~\bibnamefont{Schurtenberger}},
  \bibinfo{journal}{Nature} \textbf{\bibinfo{volume}{432}},
  \bibinfo{pages}{492} (\bibinfo{year}{2004}).

\bibitem[{\citenamefont{Saika-Voivod et~al.}(2004)\citenamefont{Saika-Voivod,
  Zaccarelli, Buldyrev, Sciortino, and Tartaglia}}]{Saika2004}
\bibinfo{author}{\bibfnamefont{I.}~\bibnamefont{Saika-Voivod}},
  \bibinfo{author}{\bibfnamefont{E.}~\bibnamefont{Zaccarelli}},
  \bibinfo{author}{\bibfnamefont{S.}~\bibnamefont{Buldyrev}},
  \bibinfo{author}{\bibfnamefont{F.}~\bibnamefont{Sciortino}},
  \bibnamefont{and}
  \bibinfo{author}{\bibfnamefont{P.}~\bibnamefont{Tartaglia}},
  \bibinfo{journal}{Phys. Rev. E} \textbf{\bibinfo{volume}{70}},
  \bibinfo{pages}{041401} (\bibinfo{year}{2004}).

\bibitem[{\citenamefont{Zaccarelli et~al.}()\citenamefont{Zaccarelli, Buldyrev,
  Nave, Moreno, Saika-Voivod, Sciortino, and Tartaglia}}]{Zaccarelli2004pre}
\bibinfo{author}{\bibfnamefont{E.}~\bibnamefont{Zaccarelli}},
  \bibinfo{author}{\bibfnamefont{S.~V.} \bibnamefont{Buldyrev}},
  \bibinfo{author}{\bibfnamefont{E.~L.} \bibnamefont{Nave}},
  \bibinfo{author}{\bibfnamefont{A.~J.} \bibnamefont{Moreno}},
  \bibinfo{author}{\bibfnamefont{I.}~\bibnamefont{Saika-Voivod}},
  \bibinfo{author}{\bibfnamefont{F.}~\bibnamefont{Sciortino}},
  \bibnamefont{and}
  \bibinfo{author}{\bibfnamefont{P.}~\bibnamefont{Tartaglia}},
  \bibinfo{note}{cond-mat/0409361}.

\bibitem[{\citenamefont{Foffi et~al.}(2004{\natexlab{b}})\citenamefont{Foffi,
  G{\"o}tze, Sciortino, Tartaglia, and Voigtmann}}]{Foffi2004}
\bibinfo{author}{\bibfnamefont{G.}~\bibnamefont{Foffi}},
  \bibinfo{author}{\bibfnamefont{W.}~\bibnamefont{G{\"o}tze}},
  \bibinfo{author}{\bibfnamefont{F.}~\bibnamefont{Sciortino}},
  \bibinfo{author}{\bibfnamefont{P.}~\bibnamefont{Tartaglia}},
  \bibnamefont{and}
  \bibinfo{author}{\bibfnamefont{T.}~\bibnamefont{Voigtmann}},
  \bibinfo{journal}{Phys.~Rev.~E} \textbf{\bibinfo{volume}{69}},
  \bibinfo{pages}{011505} (\bibinfo{year}{2004}{\natexlab{b}}).

\bibitem[{\citenamefont{Sperl}(2003{\natexlab{b}})}]{Sperl2003}
\bibinfo{author}{\bibfnamefont{M.}~\bibnamefont{Sperl}},
  \bibinfo{journal}{Phys. Rev. E} \textbf{\bibinfo{volume}{68}},
  \bibinfo{pages}{031405} (\bibinfo{year}{2003}{\natexlab{b}}).

\bibitem[{\citenamefont{Bergenholtz et~al.}(2003)\citenamefont{Bergenholtz,
  Poon, and Fuchs}}]{Bergenholtz2003}
\bibinfo{author}{\bibfnamefont{J.}~\bibnamefont{Bergenholtz}},
  \bibinfo{author}{\bibfnamefont{W.}~\bibnamefont{Poon}}, \bibnamefont{and}
  \bibinfo{author}{\bibfnamefont{M.}~\bibnamefont{Fuchs}},
  \bibinfo{journal}{Langmuir} \textbf{\bibinfo{volume}{19}},
  \bibinfo{pages}{4493} (\bibinfo{year}{2003}).

\bibitem[{\citenamefont{Gleim et~al.}(1998)\citenamefont{Gleim, Kob, and
  Binder}}]{Gleim1998}
\bibinfo{author}{\bibfnamefont{T.}~\bibnamefont{Gleim}},
  \bibinfo{author}{\bibfnamefont{W.}~\bibnamefont{Kob}}, \bibnamefont{and}
  \bibinfo{author}{\bibfnamefont{K.}~\bibnamefont{Binder}},
  \bibinfo{journal}{Phys.~Rev.~Lett.} \textbf{\bibinfo{volume}{81}},
  \bibinfo{pages}{4404} (\bibinfo{year}{1998}).

\end{thebibliography}
\bibliographystyle{apsrev}
\clearpage
\newpage

%\begin{figure}[h]
%$\begin{array}{c@{\hspace{0in}}c@{\hspace{0in}}c}
%\multicolumn{1}{l}{\mbox{\bf (a)}} &
%    \multicolumn{1}{l}{\mbox{\bf (b)}}  &
%    \multicolumn{1}{l}{\mbox{\bf (c)}}\\ [-0.cm]
%%\epsfxsize=2in
% \includegraphics[width=.25\textwidth]{../FIGURES/fig8a}
%&

%    \includegraphics[width=.25\textwidth]{../FIGURES/fig8b}
%&

%    \includegraphics[width=.25\textwidth]{../FIGURES/fig8c}

%\\ [0.4cm]
%%\mbox{ ($\mathbf\phi=0.10 \, \, \, T=0.05$)} & \mbox{ ($\mathbf\phi=0.10 \, \, \, T=0.15$)}
%\end{array}$

%\caption{
%Configurations in time for the quench at $T_f=0.05$ and $\phi=0.10$:
%(a) Newtonian Dynamics and (b) Brownian Dynamics. (c) same as (a) for
%$T_f=0.15$ and $\phi=0.10$}
%\label{fig8}
%\end{figure}

\begin{figure}[h]
\caption{
Snapshots of the configurations at different times from the quench at $T_f=0.05$ and $\phi=0.10$:
(a) Newtonian Dynamics and (b) Brownian Dynamics. (c) same as (a) for
$T_f=0.15$ and $\phi=0.10$}
\label{fig8}
\end{figure}

%\begin{figure}[h]
%$\begin{array}{c@{\hspace{.0in}}c}
%\multicolumn{1}{l}{\mbox{\bf (a)}} &
%    \multicolumn{1}{l}{\mbox{\bf (b)}} \\ [-0.cm]
%%\epsfxsize=2in
% \includegraphics[width=.25\textwidth]{../FIGURES/fig7a}
%&
%%    \epsfxsize=2in
%    \includegraphics[width=.25\textwidth]{../FIGURES/fig7b}\\ [0.4cm]
%\multicolumn{1}{l}{\mbox{\bf (c)}} &
%    \multicolumn{1}{l}{\mbox{\bf (d)}} \\ [-0.cm]
%%\epsfxsize=2in
% \includegraphics[width=.25\textwidth]{../FIGURES/fig7c}
%&
%%    \epsfxsize=2in
%    \includegraphics[width=.25\textwidth]{../FIGURES/fig7d}\\ [0.4cm]

%\mbox{ ($\mathbf T_f=0.05$)} & \mbox{ ($\mathbf T_f=0.15$)}
%\end{array}$

%\caption{
%Snapshots of the final configuration for $T_f=0.05$ (a) and $T_f=0.15$
%(b) at densities between $\phi=0.01$ and $\phi=0.10 $. \\Snapshots of
%for $T_f=0.05$ (c) and $T_f=0.15$ (d) at densities between $\phi=0.10$
%and $\phi=0.50$. The unity of length is always $\sigma_b=1$ and is
%graphically kept constant in each figure, as consequence an increase in
%density implies a smaller simulation box. }
%\label{fig7}
%\end{figure}

\begin{figure}[h]
$\begin{array}{c@{\hspace{1.2in}}c}
\mbox{ ($\mathbf T_f=0.05$)} & \mbox{ ($\mathbf \Large T_f=0.15$)}\\ \\ \\
\end{array}$

\caption{
Snapshots of the final configurations for $T_f=0.05$ (top left) and $T_f=0.15$
(top right) for $0.01 < \phi < 0.5$.The unity of length ($\sigma_b$) is
graphically kept constant for the first  and the last three rows. Hence, an increase in density implies a smaller simulation box. }
\label{fig7}
\end{figure}

\begin{figure}[tbh]
\includegraphics[width=.5\textwidth]{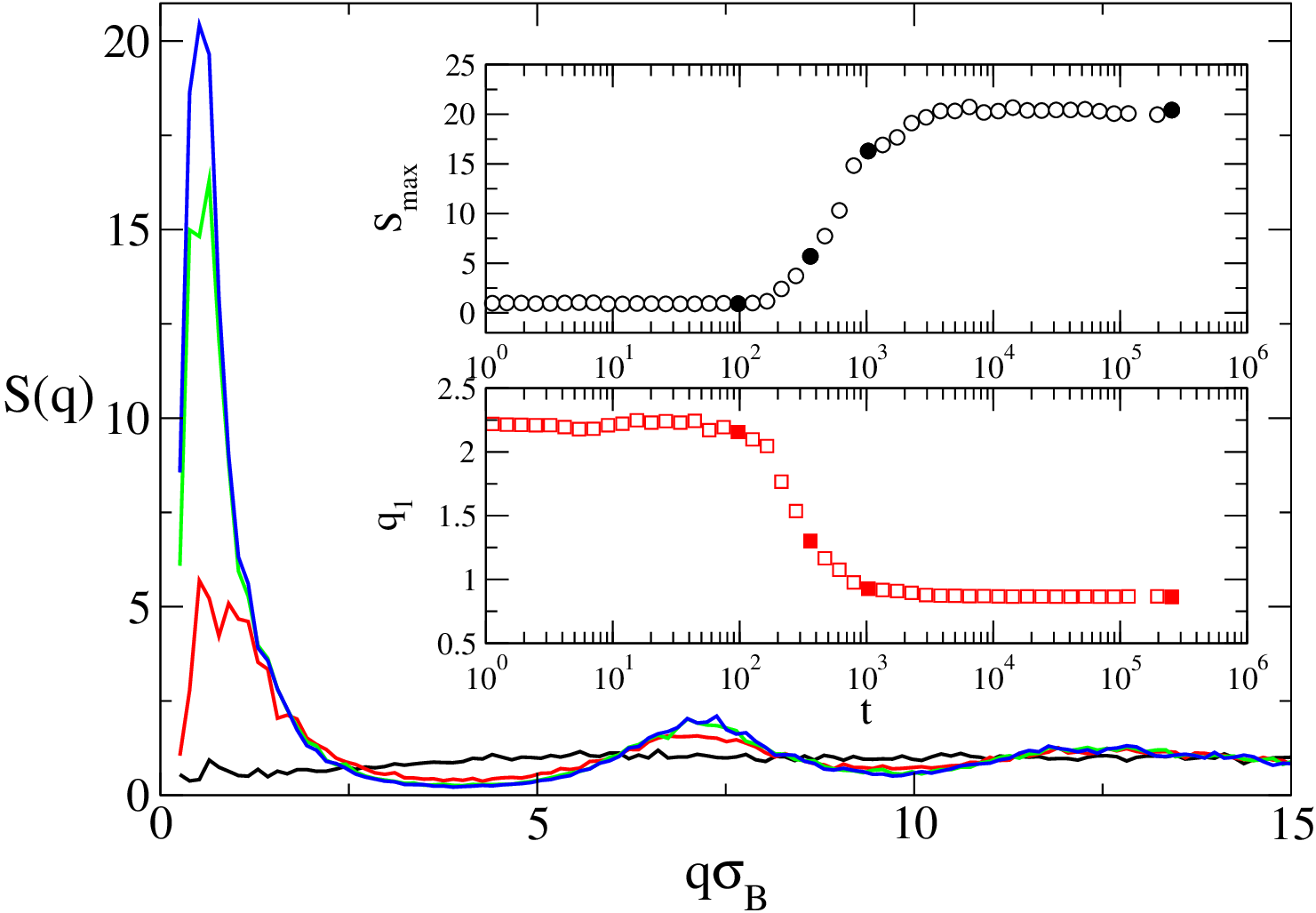}\\
\includegraphics[width=.5\textwidth]{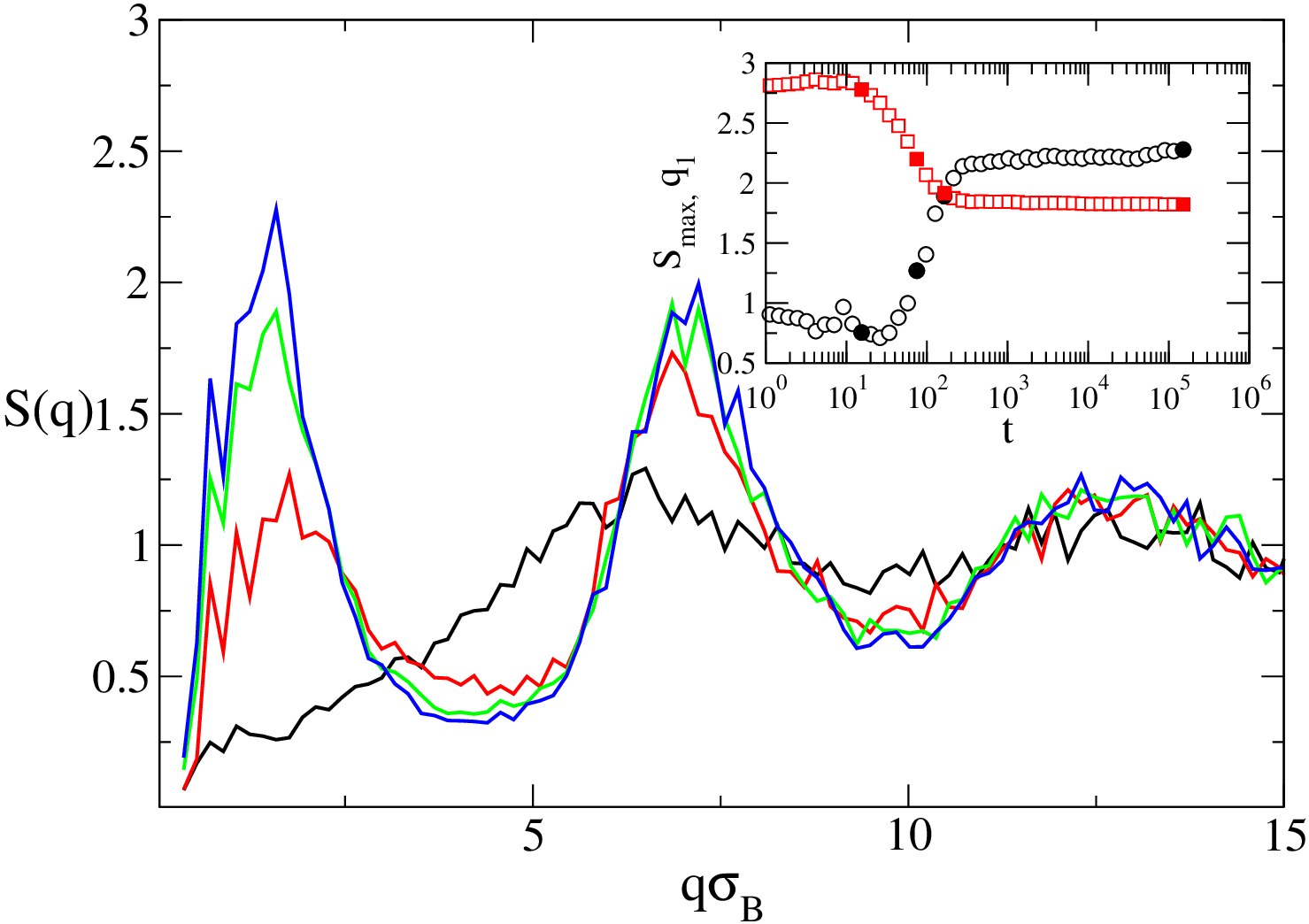}\\
\includegraphics[width=.5\textwidth]{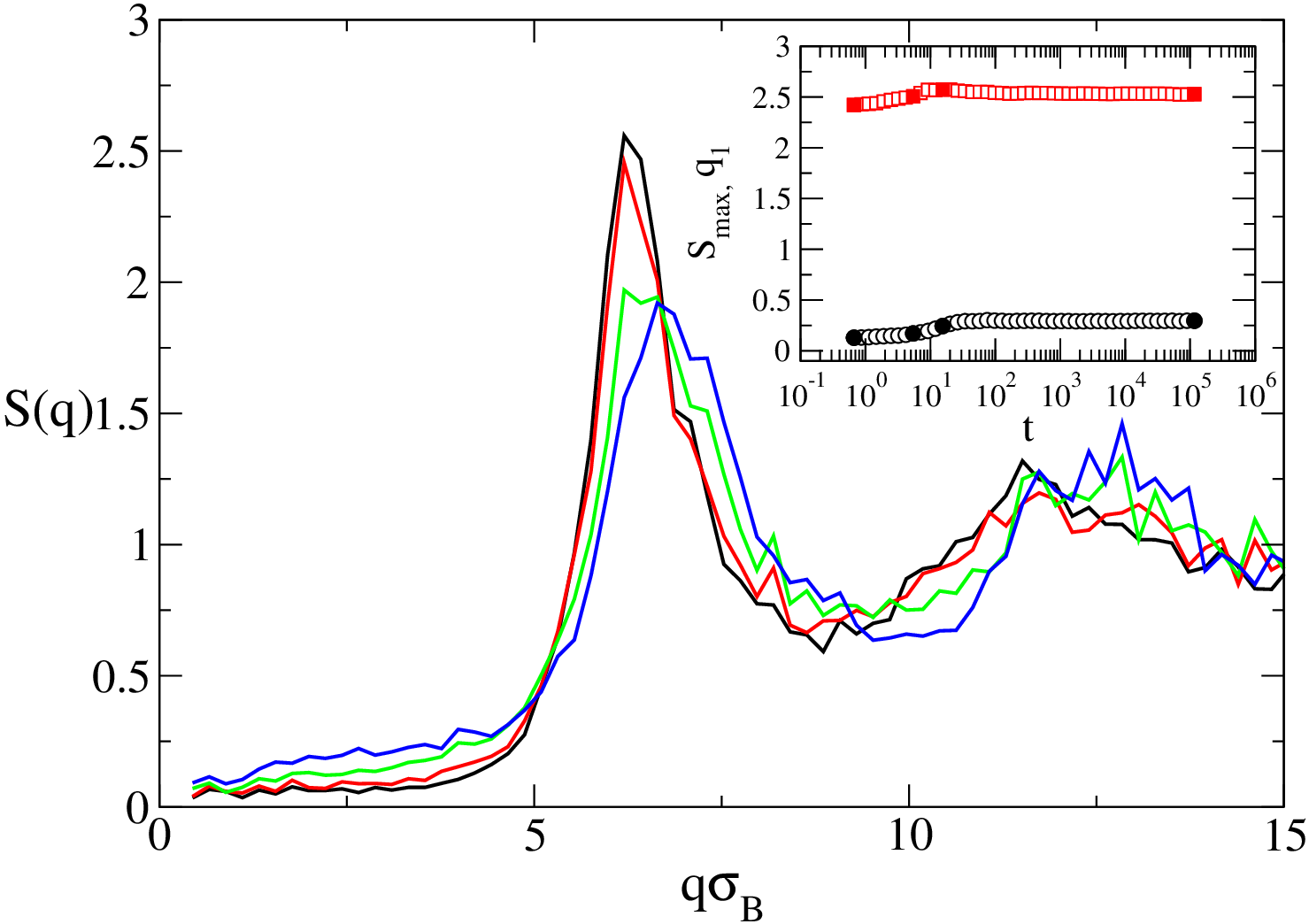}
\caption{Time evolution of the static structure factor $S(q)$  for $T_f=0.05$ at three
packing fractions: (a) $\phi=0.10$, (b) $\phi=0.25$ (the critical
packing fraction), (c) $\phi=0.50$. In the inset evolution of the
height of the first maximum and its position expressed by the quantity
$q_1$ is shown(see text for details). The full symbols in the inset represent the times at
which the structure factors are shown in the main figure.}
\label{fig3}
\end{figure}

\begin{figure}[tbh]
\includegraphics[width=.5\textwidth]{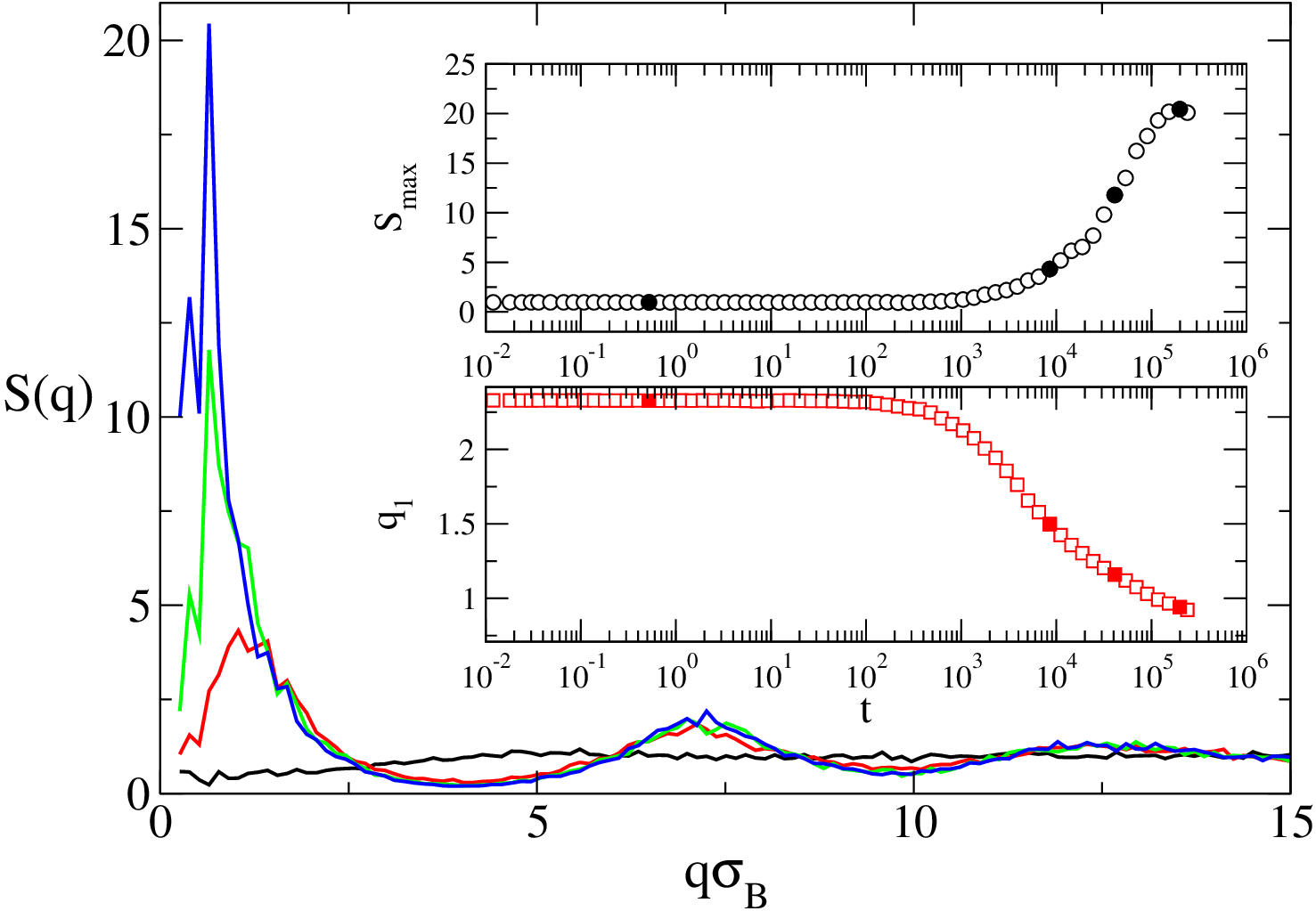}\\
\includegraphics[width=.5\textwidth]{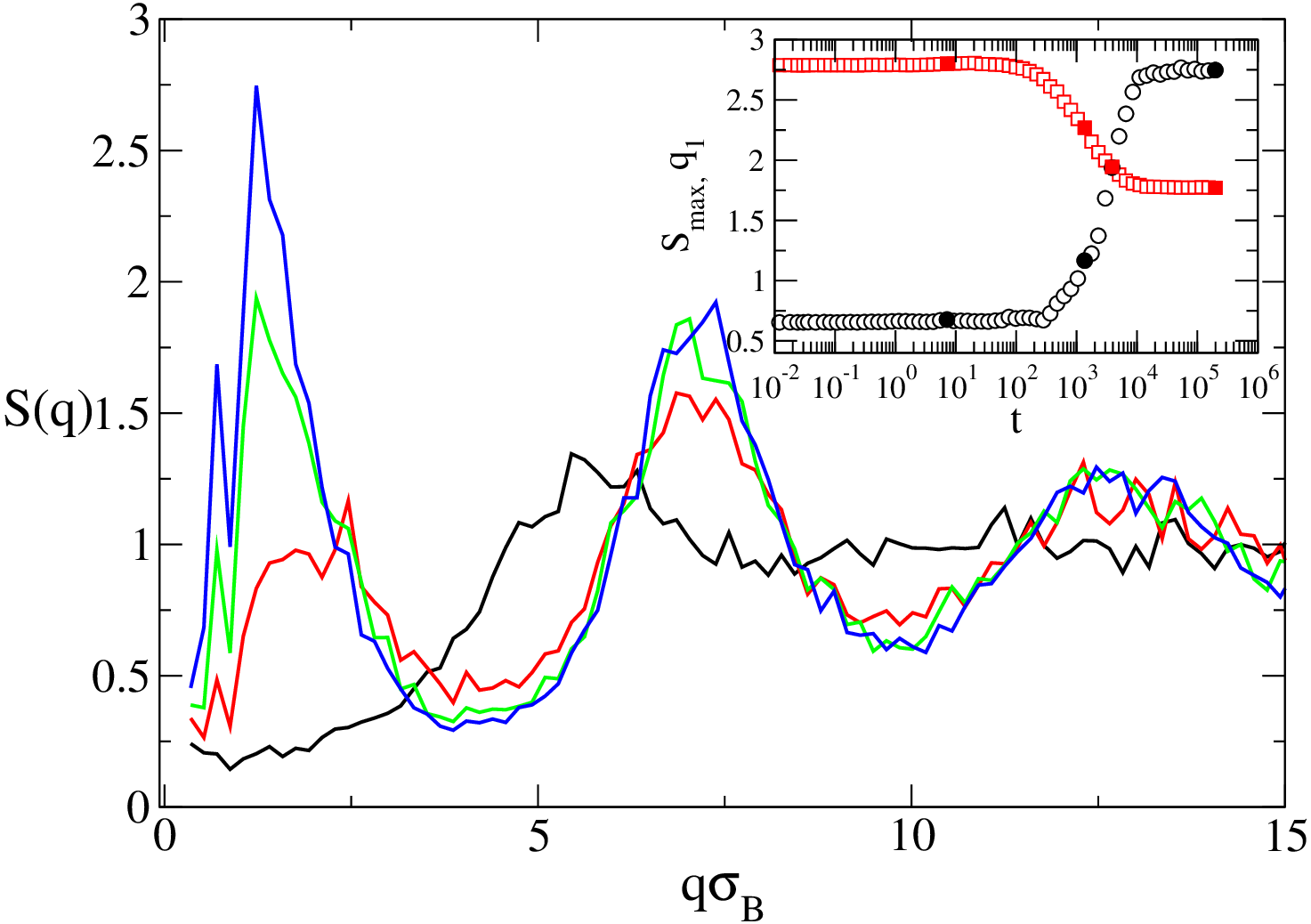}\\
\includegraphics[width=.5\textwidth]{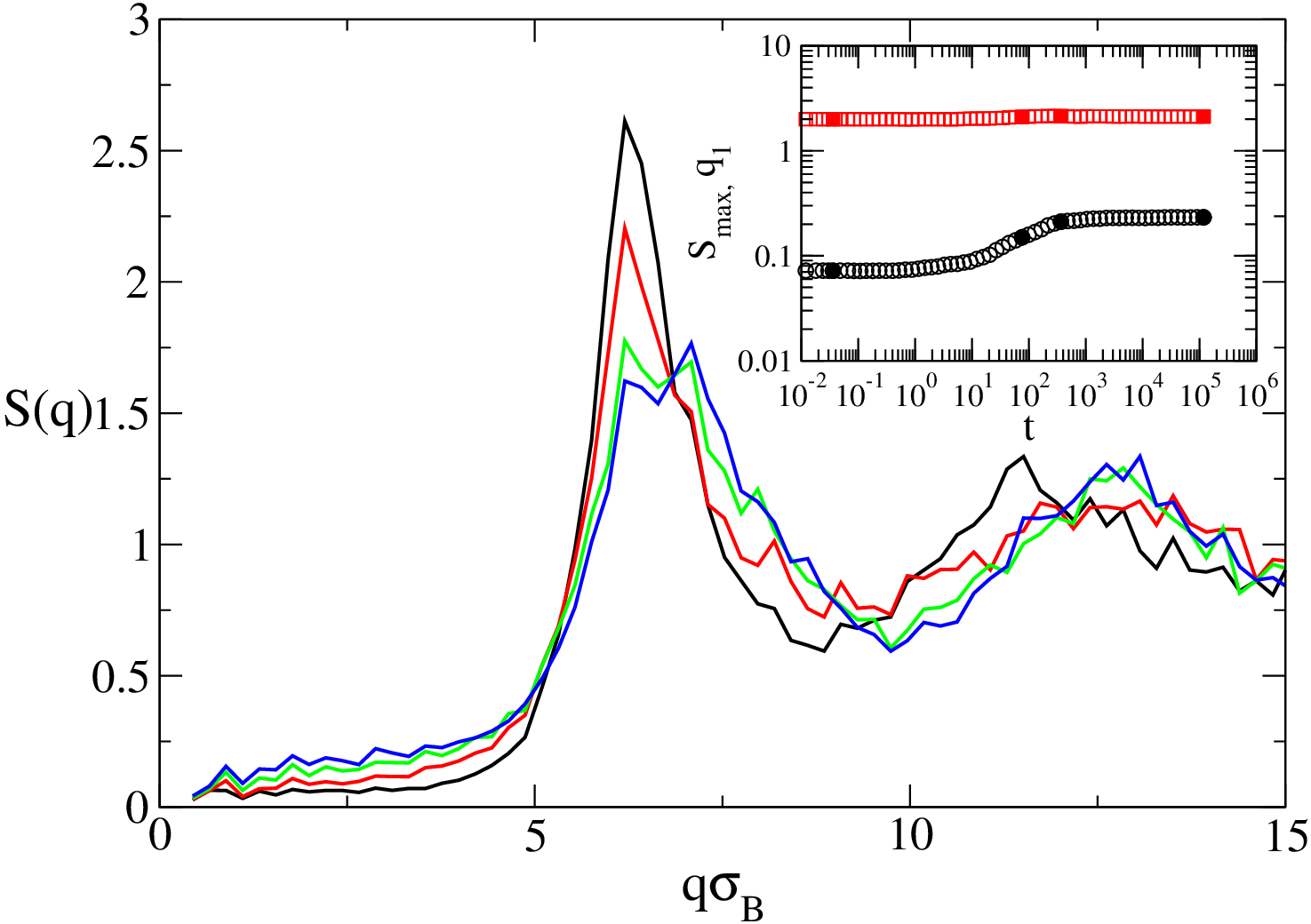}
\caption{Same as Fig.\ref{fig3} for Brownian Dynamics.}
\label{fig4}
\end{figure}

\begin{figure}[tbh]
\includegraphics[width=.5\textwidth]{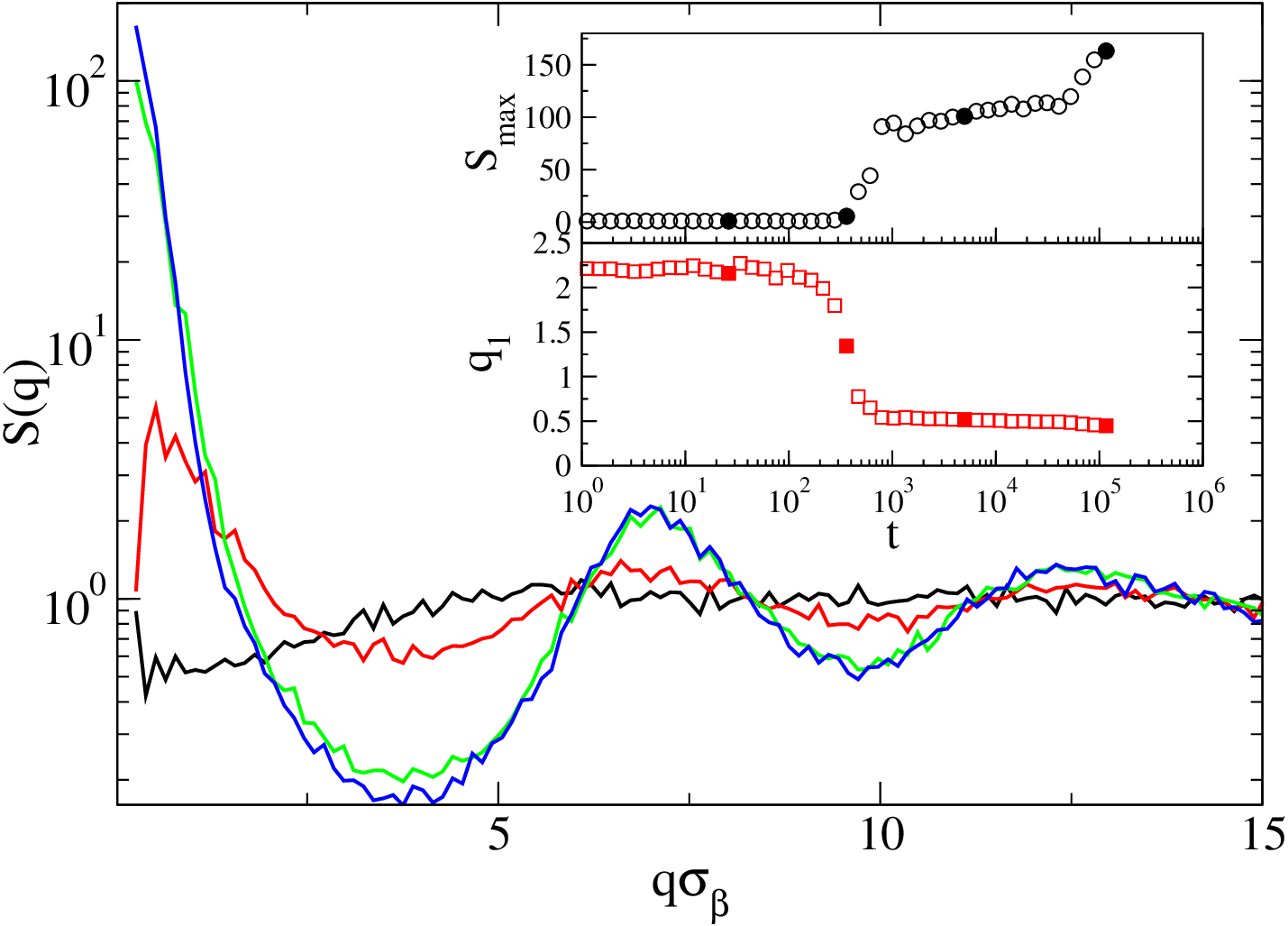}\\
\includegraphics[width=.5\textwidth]{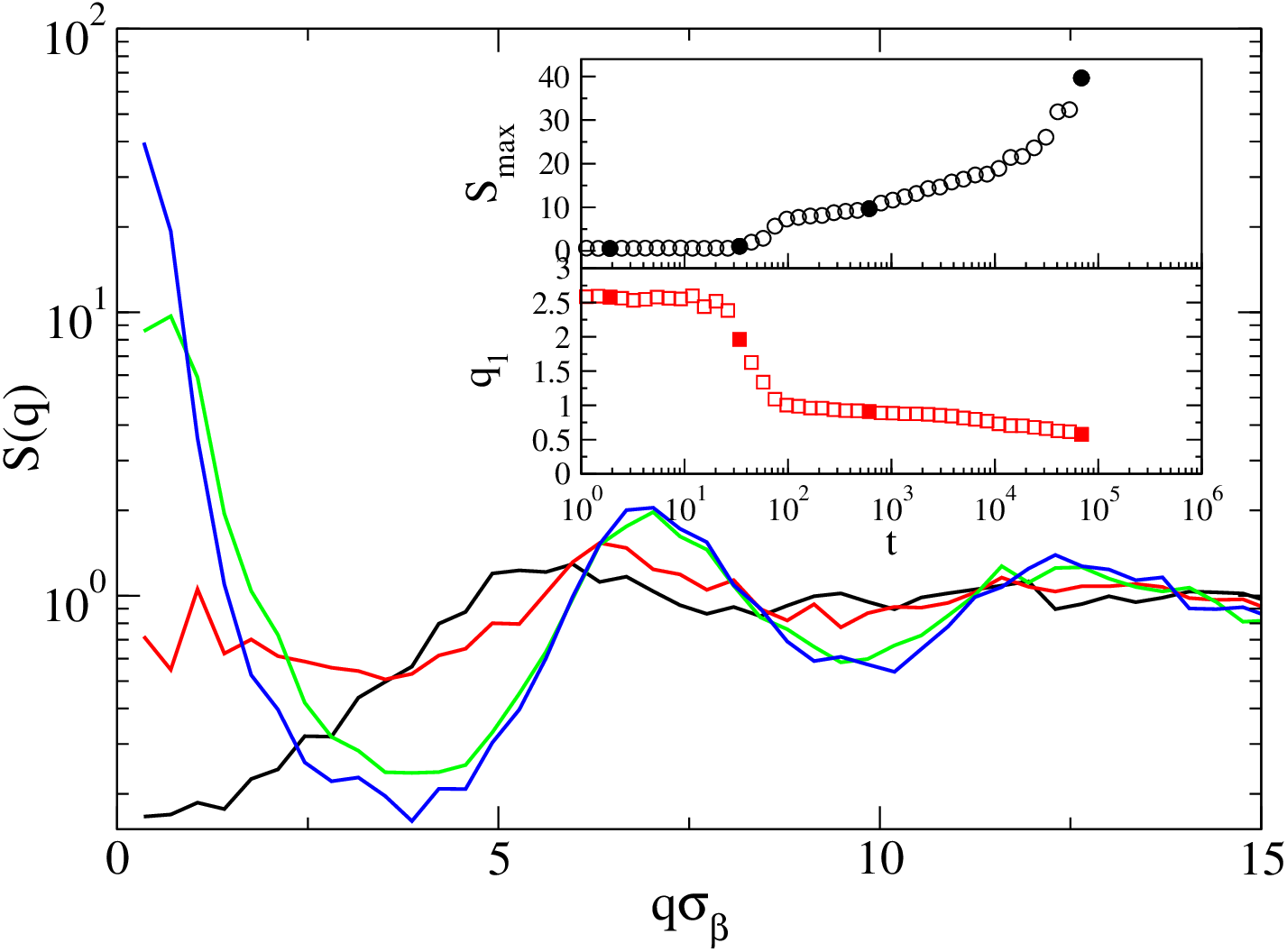}\\
\includegraphics[width=.5\textwidth]{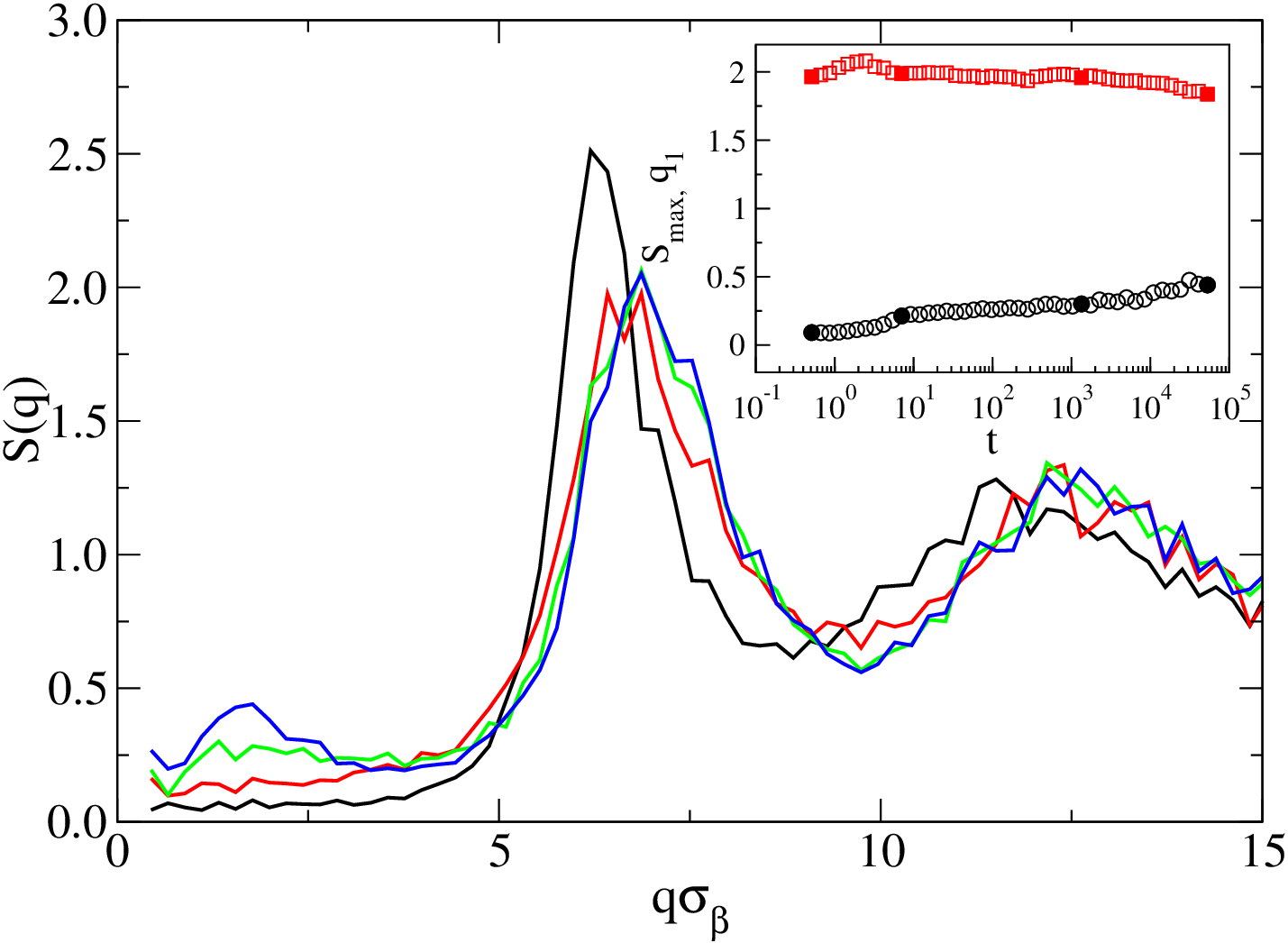}
\caption{Same as Fig.\ref{fig3} for $T_f=0.15$.}
\label{fig5}
\end{figure}

\begin{figure}[tbh]
\includegraphics[width=.5\textwidth]{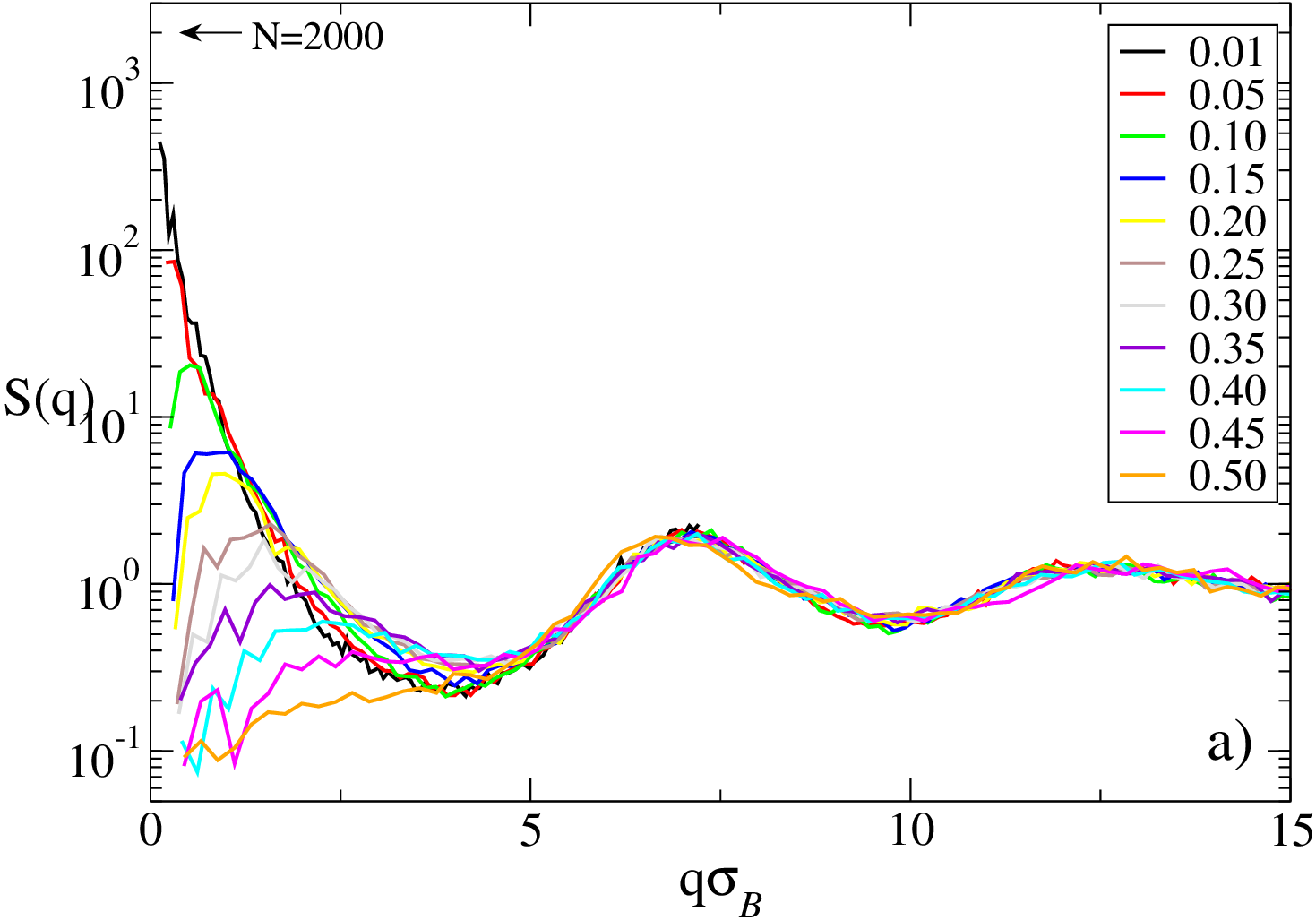}\\
\includegraphics[width=.5\textwidth]{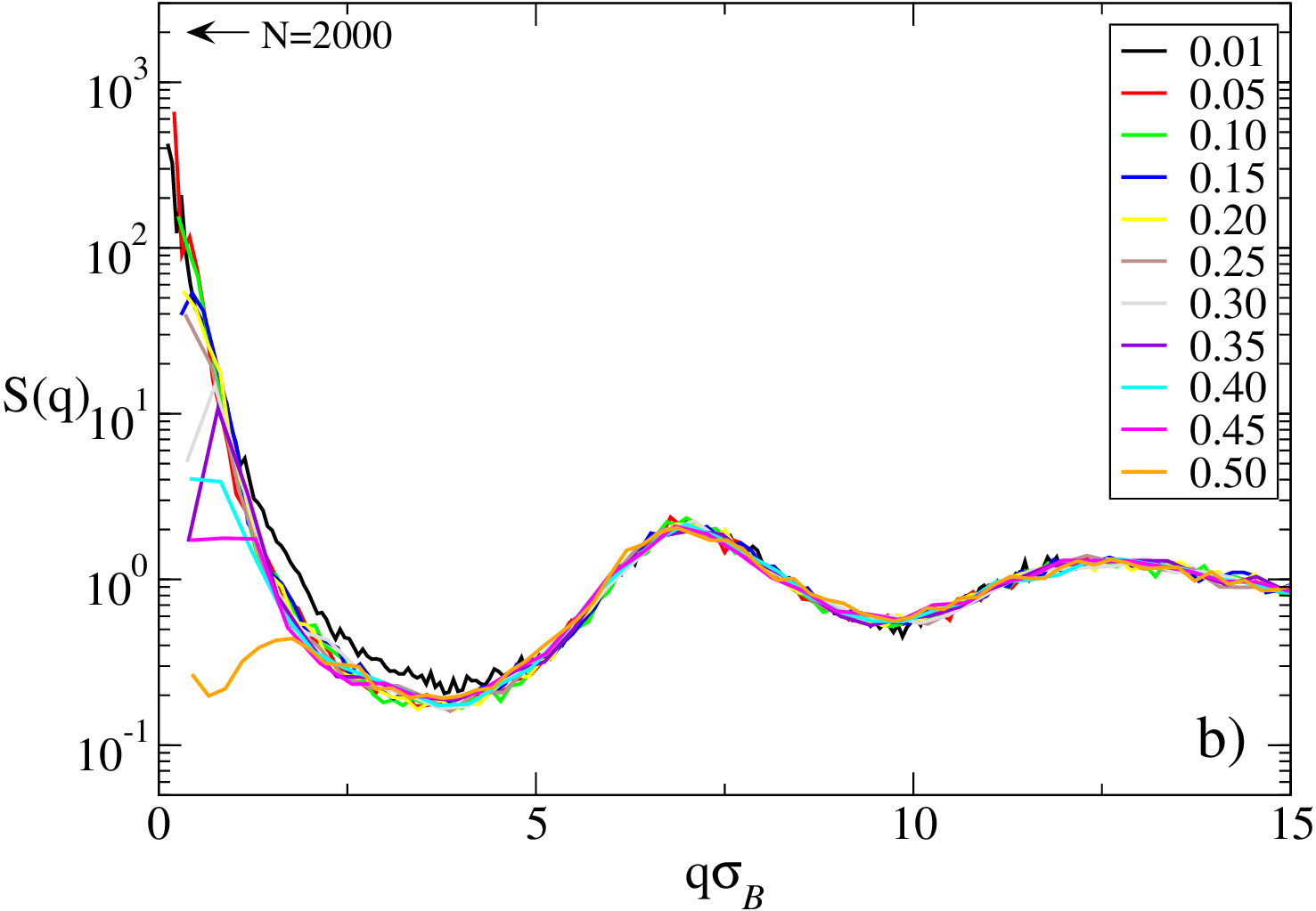}
\caption{Static structure factors $S(q)$ of the final configuration for different packing fractions. (a) $T_f=0.05$, (b) $T_f=0.15$}
\label{fig6}
\end{figure}

\begin{figure}[tbh]
\caption{Representation of the studied two dimensional slabs. The
three slabs are chosen at an height $z=2.4$, $z=12.0$ and $22.6$ and
they have a width of $4 \sigma_b$}
\label{fig9}
\end{figure}

\begin{figure}[h]
%$\begin{array}{c@{\hspace{.0in}}c}
%\multicolumn{1}{l}{\mbox{\bf (a)}} &
%    \multicolumn{1}{l}{\mbox{\bf (b)}} \\ [-0.cm]
%\epsfxsize=2in
% \includegraphics[width=.25\textwidth]{fig10a}
%&
%%    \epsfxsize=2in
%    \includegraphics[width=.25\textwidth]{fig10b}\\ [0.4cm]

%\mbox{ ($\mathbf T_f=0.05$)} & \mbox{ ($\mathbf T_f=0.15$)}
%\end{array}$

\caption{Snapshots of the two dimensional slabs (see Fig.\ref{fig9}) of the final configurations at three values of the $z$-coordinate. (a) $T_f=0.05$ and $\phi=0.10$; (b)  $T_f=0.15$ and $\phi=0.10$}
\label{fig10}
\end{figure}

\begin{figure}[tbh]
\includegraphics[width=.5\textwidth]{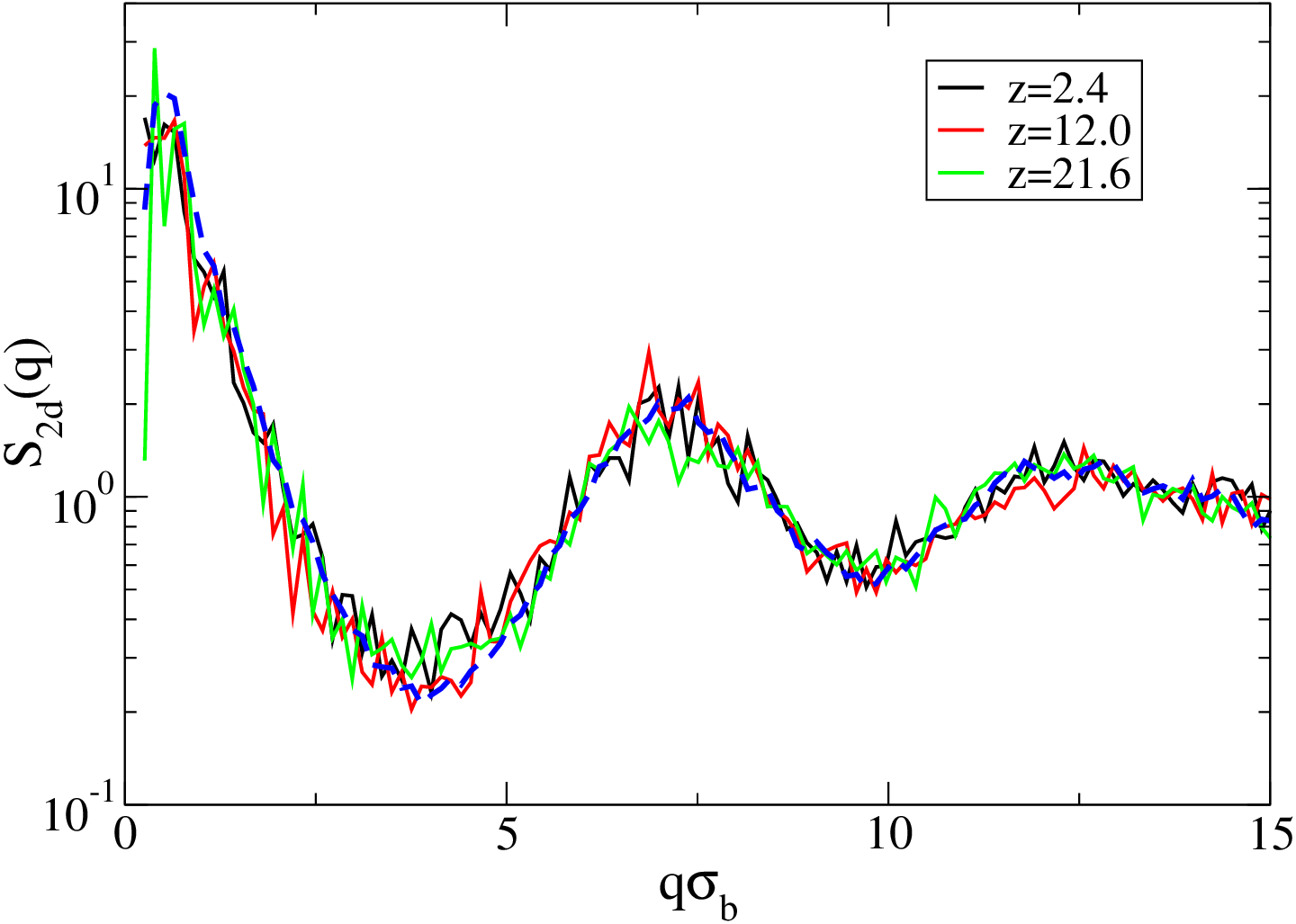}\\
\includegraphics[width=.5\textwidth]{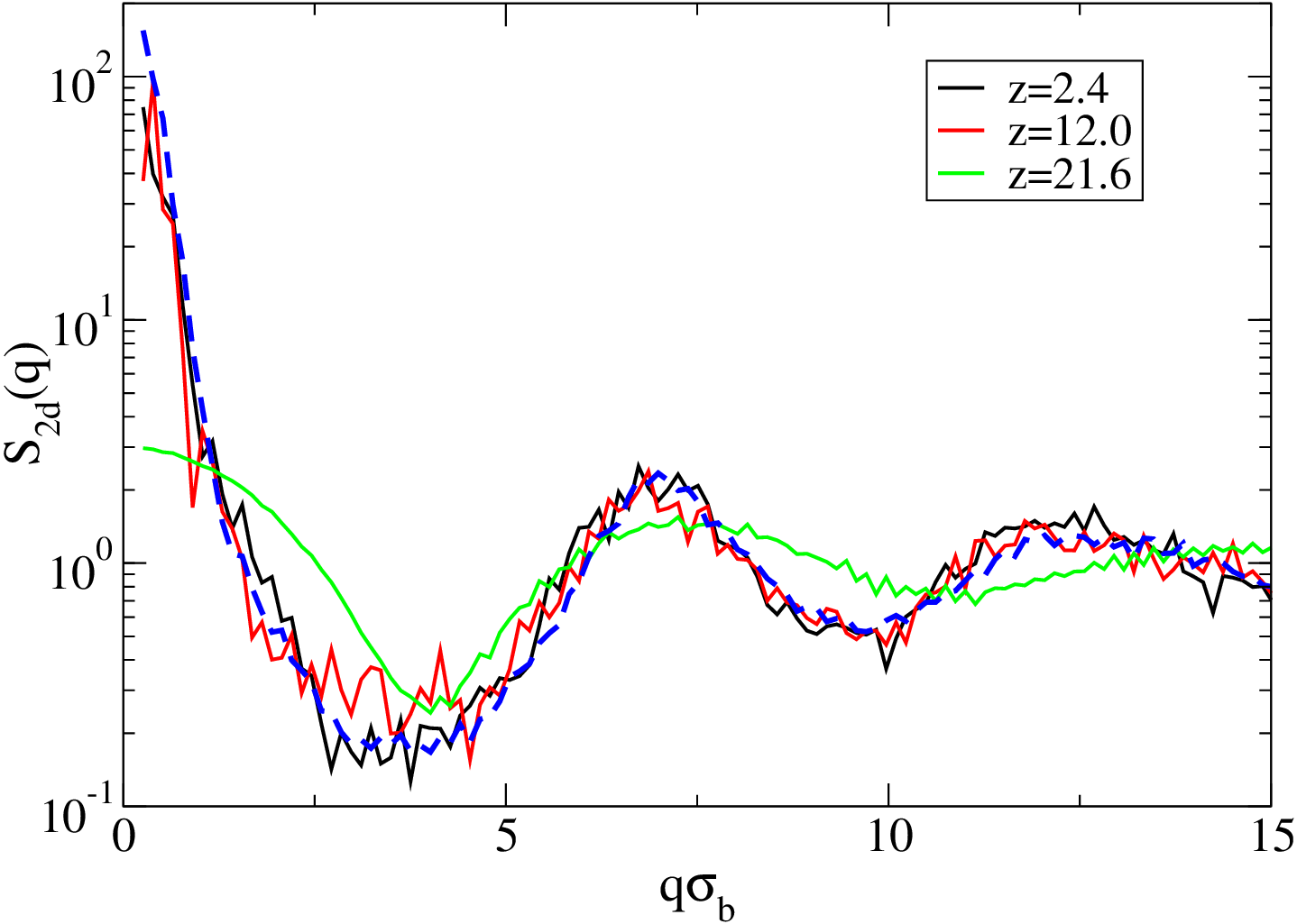}
\caption{
    Two dimensional structure factor $S_{2d}(q)$ for different
    values of the $z$-coordinate (see text for details). (a)
    $T_f=0.05$ and $\phi=0.10$; (b) $T_f=0.15$ and
    $\phi=0.10$. The dashed line corresponds to the three dimensional spherically averaged $S(q)$.}
\label{fig11}
\end{figure}

\begin{figure}[tbh]
\includegraphics[width=.4925\textwidth]{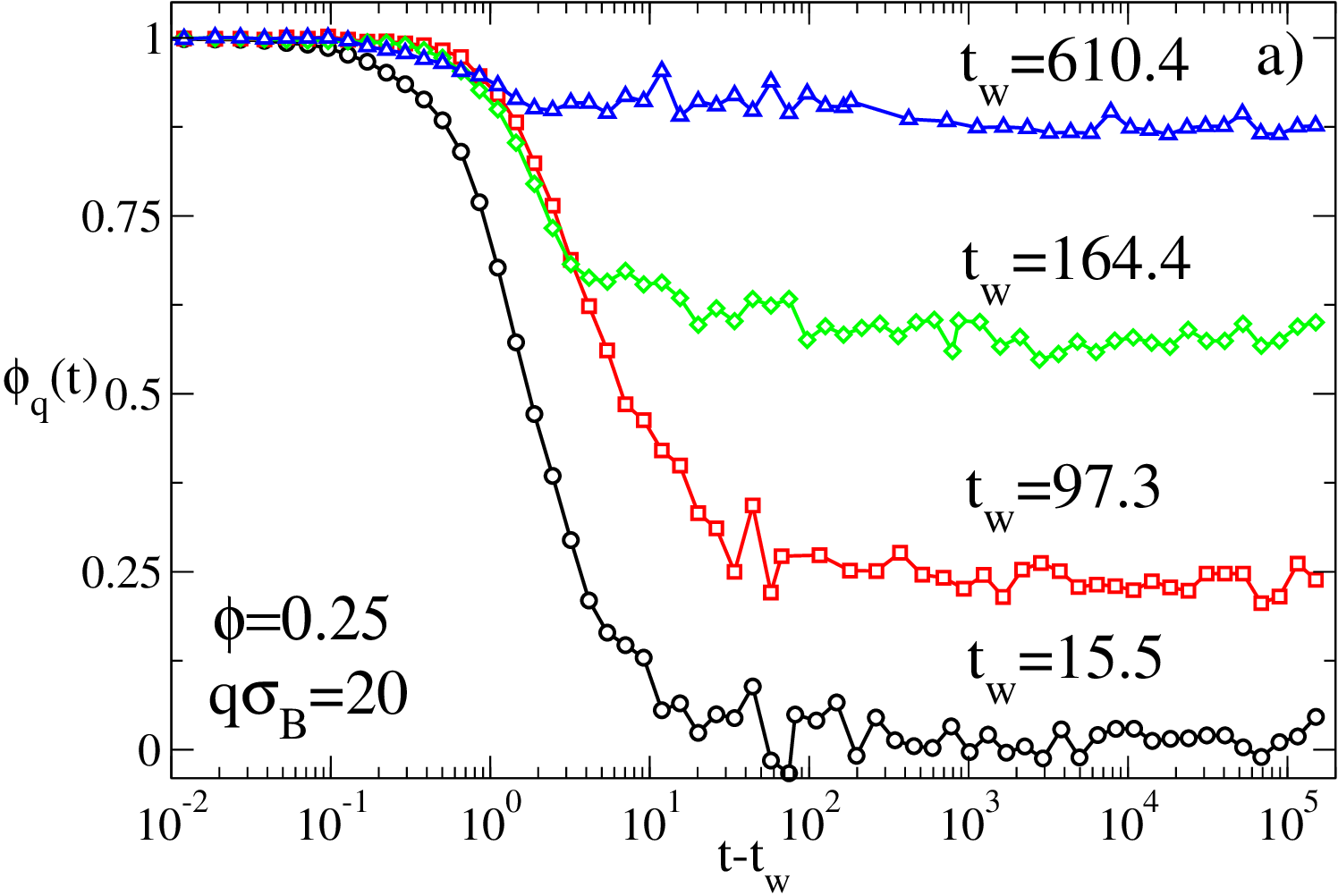}
\includegraphics[width=.5\textwidth]{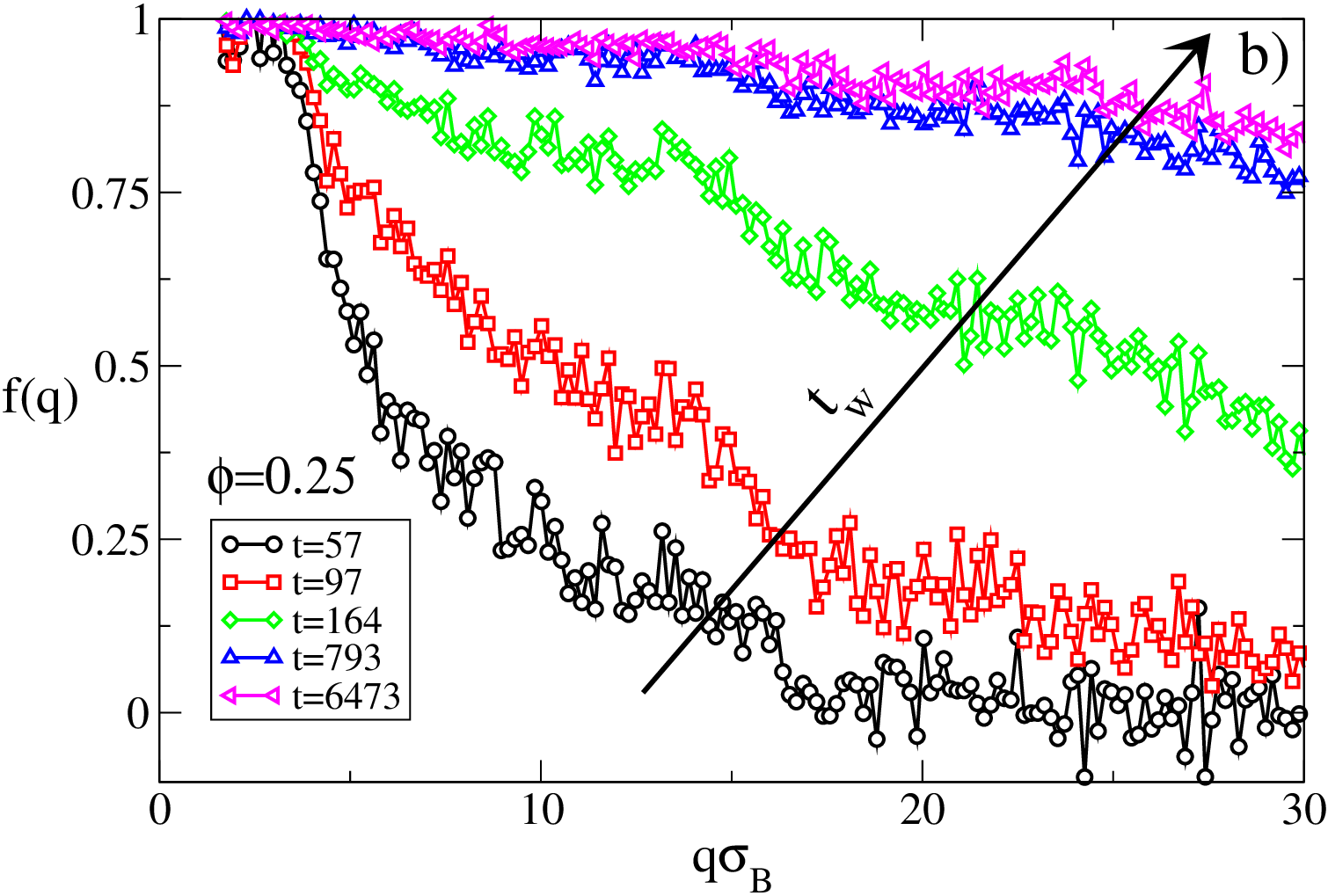}
\caption{
    a) Density-density correlation function $\phi_q(t)$ for different waiting
    times $t_w$ after the quench. This case corresponds to $\phi=0.25$  and $T_f=0.05$.
    Correlation functions  are calculated for $q\sigma_B=20$. (b) Time evolution of the non ergodicity parameter $f_q$  with increasing $t_w$ for the same case.
    }
\label{fig12}
\end{figure}

\begin{figure}[tbh]
\includegraphics[width=.5\textwidth]{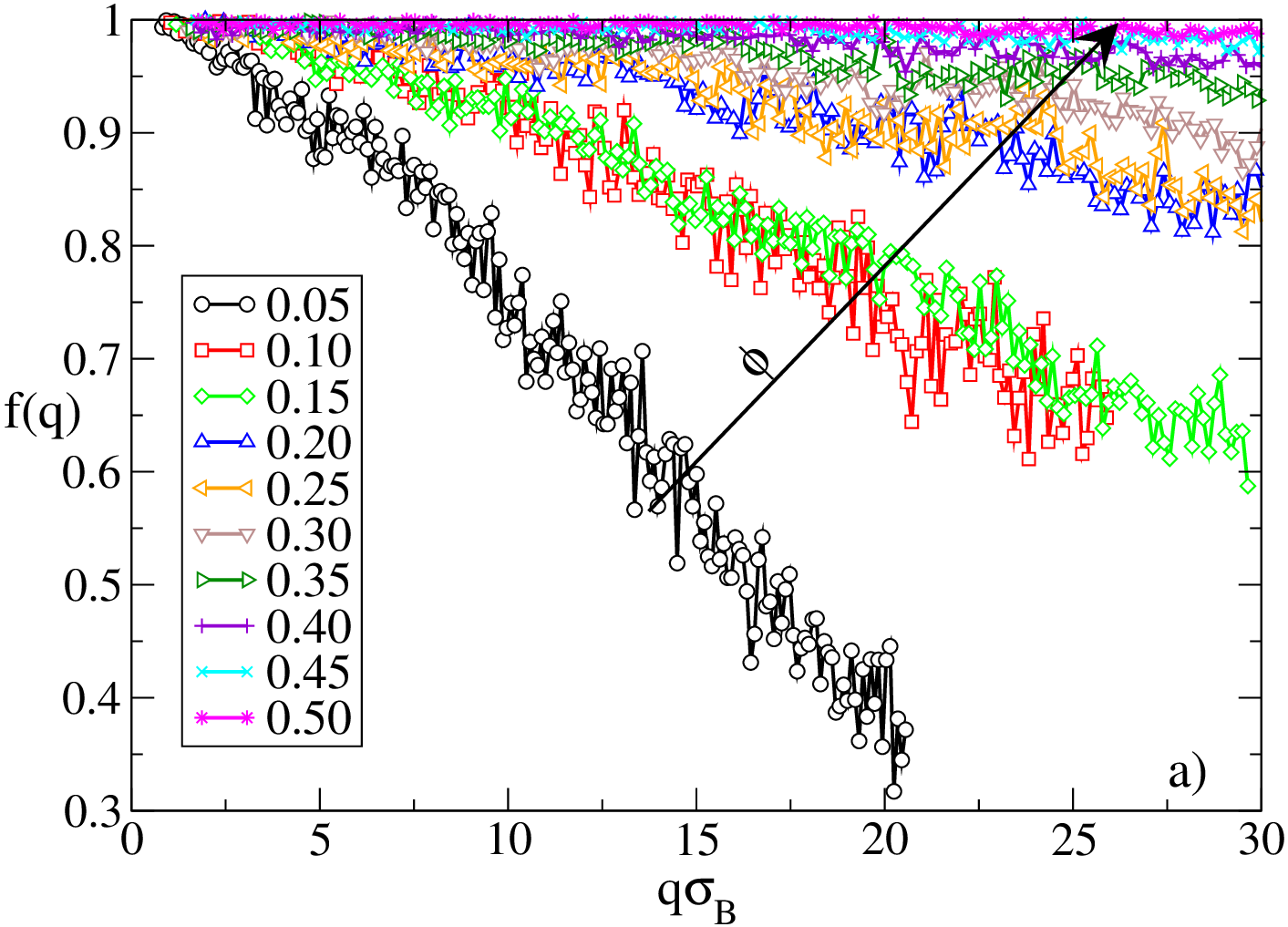}
\includegraphics[width=.5\textwidth]{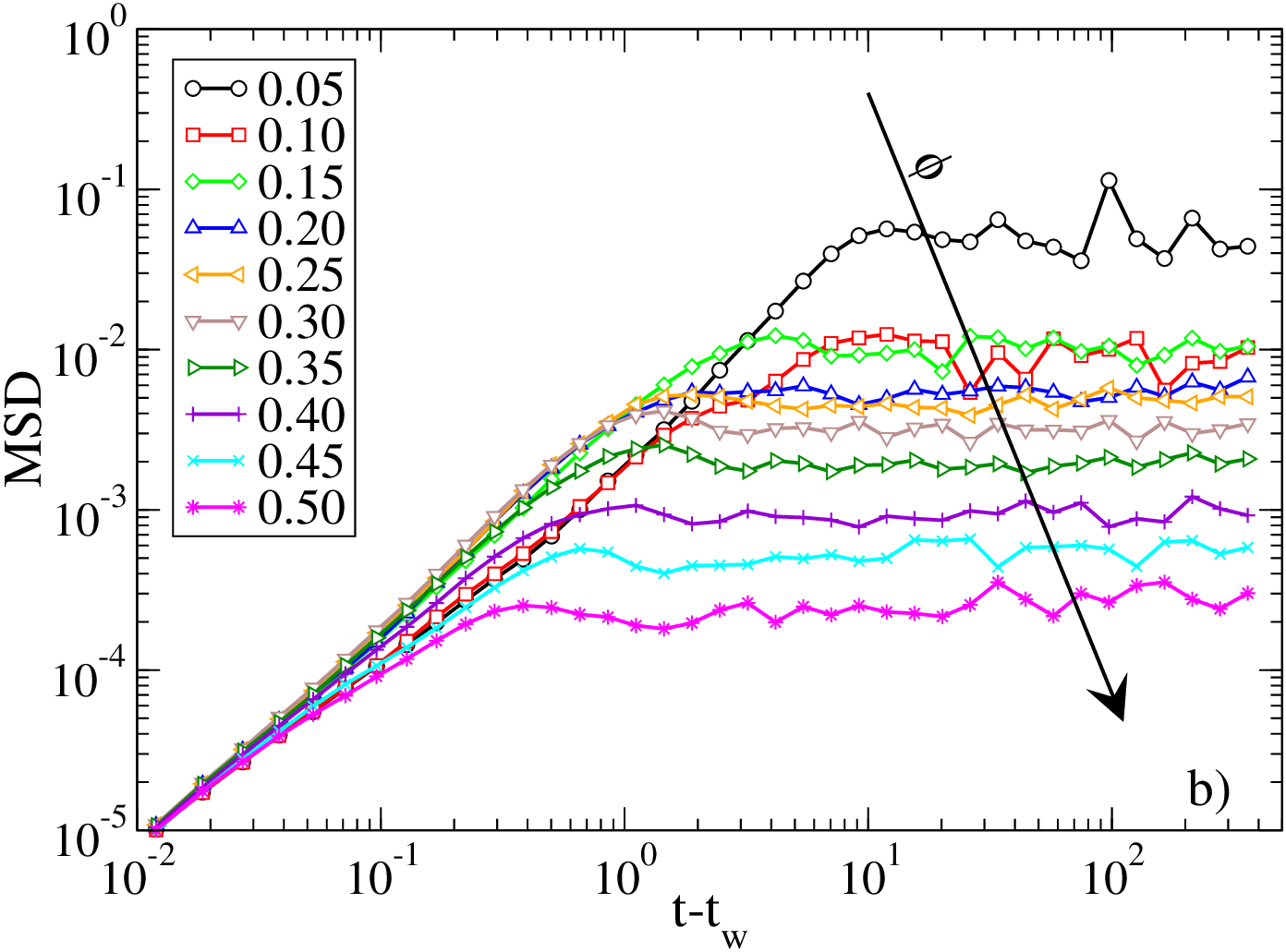}
\caption{
    a) Non ergodicity parameter for different packing fraction in the final arrested state, at $T_f=0.05$.\\ b) Mean square
    displacement  in the final arrested state for different values of $\phi$.}
\label{fig13}

\end{figure} 

\end{document}